\documentclass[aps,pra,showpacs,twocolumn,amsmath,amssymb,superscriptaddress, footinbib]{revtex4}

\usepackage[english]{babel}

\usepackage{latexsym}
\usepackage{graphicx}
\usepackage{subfigure}
\usepackage{epsfig}
\usepackage{amsfonts}
\usepackage{amssymb}
\usepackage{amsmath}
\usepackage{bm, bbm}
\usepackage{braket}
\usepackage{wrapfig}
\usepackage{hyperref}
\usepackage{tikz}
\usetikzlibrary{arrows,chains,matrix,positioning,scopes}
\makeatletter
\tikzset{join/.code=\tikzset{after node path={%
\ifx\tikzchainprevious\pgfutil@empty\else(\tikzchainprevious)%
edge[every join]#1(\tikzchaincurrent)\fi}}}
\makeatother
\tikzset{>=stealth',every on chain/.append style={join},
         every join/.style={->}}
\tikzstyle{labeled}=[execute at begin node=$\scriptstyle,
   execute at end node=$]
\usetikzlibrary{calc,arrows,shapes,snakes,automata}
\newcommand*\rows{5}

\usepackage{lipsum} % dummy text for the MWE
\usepackage[draft]{todonotes}   % notes showed
\begin{document}
\title{A Fock state lattice approach to quantum optics}

\author{Pil Saugmann and Jonas Larson}
\affiliation{Department of Physics,
Stockholm University, AlbaNova University Center, 106 91 Stockholm,
Sweden}

\date{\today}

\begin{abstract}
We analyze a set of models frequently appearing in quantum optical settings by expressing their Hamiltonians in terms of Fock-state lattices (FSLs). The few degrees-of-freedom of such models, together with the system symmetries, make the emerging FSLs rather simple such that they can be linked to known lattice models from the condensed matter community. This sheds new light on known quantum optical systems. While we provide a rather long list of models and their corresponding FSLs, we pick a few ones in order to demonstrate the strength of the method. The three-mode boson model, for example, is shown to display a fractal spectrum, and chiral evolution in the FSL characterized by localized distributions traversing along symmetric trajectories. In a second example we consider the central spin model which generates a FSL reminiscent of the Su-Schrieffer–Heeger model hosting topological edge states. We further demonstrate how the phenomena of flat bands in lattice models can manifest in related FSLs, which can be linked to so called dark states. 
\end{abstract}

\maketitle

\section{Introduction}
In physics, lattice models prove a very powerful tool for the investigation of condensed matter systems~\cite{altland}. They arise naturally in many situations, most notably in crystal structures, and at the same time they provide a direct link to computational physics as all continuous models turn into lattice models when discretized. In the quantum realm, lattice models are in one way or another a key ingredient in many suggestions for realization of quantum simulators~\cite{mb}. While some quantum single particle lattice models can be solved analytically, others can effectively be simulated on a classical computer. However, they quickly grow in complexity when more particles are added to the system even when they are non interacting, no longer allowing for an analytical solution to be found nor even being numerically simulated. Indeed, it was recently shown how quadratic bosonic lattice models with only a few degrees-of-freedom can become computationally hard to simulate, and furthermore how such models can be mapped to interacting many-body systems~\cite{FSsimulation}. Thus, alternative physical realizations of the lattice models are in place. Here, the power of lattice models becomes particularly visible, as many properties are not linked to details of the specific physical realization of the lattice model, but instead linked to the underlying symmetries of the lattice model -- the idea of universality. As long as these stay the same, so do the universal features. Keeping this in mind, it is of course interesting to look for novel ways to realize both well known and new lattice models.  

Traditionally, the geometry of the lattice is rooted in the system's spatial dimensions, i.e. we consider lattices in three or lower dimensions. In mathematics there are no such limitations and we may even have a non-integer number of dimensions. Synthetic lattice dimensions occur in various proposals for quantum simulators as a route towards new exotic models ~\cite{synt,synt1,synt2,synt3}. Examples on this are utilizing atomic internal degrees-of-freedom ~\cite{synt}, or vibrational states of the harmonic trapping potential~\cite{synt2}, which can be employed to create synthetic dimensions for ultracold atoms loaded in optical lattices. More generally, if we give up the idea of a lattice living in real space, any Hamiltonian $\hat H$, single-particle or many-body, can be described as a single-particle system if mapped into state space. In such a view, a lattice-like structure emerges by identifying the different basis states as the lattice sites. Then, $\hat H$ is a matrix in which the diagonal terms serve as on-site energies in the lattice, and the off-diagonal elements give the tunneling rates between these sites. Such a lattice will rapidly grow in complexity as the number of degrees-of-freedom is increased, and it might well be that then not much new insight is gained by such a viewpoint. Furthermore, for practical reasons, we require that the chosen basis should be physically relevant. Paradigmatic light-matter models found in quantum optics seem perfect for this; it is possible to isolate a few relevant degrees-of-freedom, and the natural basis comprises the bare Fock states. Hence, their Hamiltonians give rise to Fock state lattices (FSLs). The FSL emerging from the  Jaynes-Cummings (JC) model was already introduced in Ref.~\cite{FSL1}, and studied further in Ref.~\cite{jctop} to explore a new type of topological matter formed by quantized light. We note that, historically, similar lattice ideas were discussed in terms of so-called Glauber-Fock lattices in wave-guide systems of classical light~\cite{synt3}.

In this work we bring forward a plethora of light-matter type models, even though we shall focus on a few of them. It turns out that these capture the general ideas, at the same time as the physics is extremely rich despite their simplicity. Taking the multi-mode JC model as an example, in the high detuned limit the spin degree-of-freedom freezes out and one is left with an effective quadratic Hamiltonian consisting only of bosonic degrees-of-freedom. The Fock states provide a natural basis to study such systems, and the sparseness of the Hamiltonian makes the FSLs convenient for getting insight into the models. For three bosonic modes, a triangular FSL emerges, and by further imposing a synthetic magnetic flux it is shown how the system's spectrum becomes fractal, akin to the Hofstadter butterfly spectrum of two dimensional (2D) tight-binding lattice models exposed to homogeneous perpendicular magnetic fluxes~\cite{hof}. It should be noted, though, that the FSLs studied in this paper lack translational invariance, and hence the fractal structure is not deriving from the same mechanism as for the Hofstadter models. The properties of our fractal spectrum imply that initial Fock states will follow very symmetric curves within the FSL. As another example we consider the central spin model in which a single spin-1/2 particle interacts identically to $N$ other spin-1/2 particles. By tuning the interaction it is possible to create a FSL bearing similarities to the Su-Schrieffer–Heeger (SSH) model, and like for the SSH model one finds exponentially localized edge states. Related to this we also discuss a few models which support a large number of degenerate $E=0$ eiegenstates, which serve as the counterpart of flat bands in for example Lieb lattices. 

The structure of the paper is the following. In the next section we systematically introduce the FSLs. First for the resonant JC models, and from there we elaborate on how one may design interesting FSLs by looking at related models. In Sec.~\ref{sec:symmetry} we discuss what role the underlying symmetries has for the lattice dimensions.  Multi-mode JC models in the large detuned limit, and in the presence of a synthetic flux, are considered in  Sec.~\ref{sec:detuned}. For these systems, we see how a fractal like structure of energy spectrum emerges for certain fluxes, which motivates us to in Sec.~\ref{sec:evolution} study how initially localized states evolve within the FSL.  Section~\ref{sec:ssh} discusses a connection between the central spin model and the SSH model, namely the appearance of exponentially localized edge sates in the FSL. Finally, in Sec.~\ref{sec:conclusion} we summarize our findings.
 
\section{Resonant Jaynes-Cummings models and their Fock state lattices}\label{sec:resonant}
\subsection{Multi-mode Jaynes-Cummings models}
In this section we introduce the JC model, which, as the simplest example of a spin-boson model, will serve as a starting point for building other models with interesting underlying FSLs. The JC model, first introduced by Edwin Jaynes and Fred Cummings in 1963~\cite{JC}, provides a fully quantum mechanical treatment of an atom interacting with an electromagnetic field~\cite{Themis}. The original model consists of a two-level atom, pseudo spin-1/2 particle, interacting with a single bosonic mode representing the electromagnetic field. The interaction can be understood as an exchange of excitations between the atom and the field; when the ground state is excited to the excited state, the number of bosons is lowered by one, and vice versa for de-excitation to the ground state. This gives rise to a Hamiltonian on the form
\begin{equation}
    \hat{H}_\mathrm{JC}=\hat{H}_{0}+\hat{H}_\mathrm{int},
\end{equation}
with the non-interacting, or bare, part of the Hamiltonian $\hat{H}_{0}$ comprises the free energies of the field and atom, i.e.
\begin{equation}
    \hat{H}_{0}=\omega\hat{n}+\frac{\Omega}{2}\hat{\sigma^z},
\end{equation}
where we have set $\hbar=1$, and $\omega$ is the photon frequency and $\Omega$ the transition frequency for the two-level atom. Further, for the boson number operator $\hat n|n\rangle=n|n\rangle$ for some Fock state $|n\rangle$, and $\hat\sigma^z$ is the Pauli $z$-matrix which upon acting on the bare atomic states; $\hat\sigma^z|g\rangle=-|g\rangle$ and $\hat\sigma^z|e\rangle=|e\rangle$, with $|g\rangle$ and $|e\rangle$ the lower and upper atomic states respectively. Throughout, following the logic of bosonic states, we will also call the atomic basis states, like $\ket{g}$ and $\ket{e}$, as Fock states. Returning to the interaction part $\hat H_\mathrm{int}$, both the atom and the bosonic mode emit and absorb excitations, and we may write the interaction Hamiltonian as~\cite{Themis}
\begin{equation}
    \hat{H}_\mathrm{int}= g\left(\hat{\sigma}^+ +\hat{\sigma}^-\right)\left(\hat{a}+\hat{a}^{\dagger}\right),\label{eq:RWA}
\end{equation}
with $g$ the light-matter interaction strength. It includes four different contributions; two of which preserve the number of excitations in the system, and two corresponding to simultaneous excitation/de-excitation of the two degrees-of-freedom. The latter do not conserve the bare energy of the system, and for now we shall drop them both. This is what is called the rotating wave approximation (RWA) and it leads to an interaction on the form~\cite{Themis}
\begin{equation}\label{jcintham}
    \hat{H}_\mathrm{int}=g\left(\hat{a}^\dagger\hat{\sigma}^{-}+\hat{\sigma}^{+}\hat{a}\right).
\end{equation}
Moving to the interaction picture with respect to $\omega\hat N$, with $\hat N=\hat n+\frac{1}{2}\hat\sigma^z$ the excitation number operator, the number of free parameters can be reduced from three to two. Defining the atom-field detuning as $\Delta=\Omega-\omega$, the Hamiltonian (in this interaction picture) can be written as
\begin{equation}
\hat{H}_\mathrm{JC}= \frac{\Delta}{2}\hat{\sigma}^z+g\left(\hat{a}^\dagger\hat{\sigma}^{-}+\hat{\sigma}^{+}\hat{a}\right),\label{JC-Ham}   
\end{equation}
which is the celebrated JC Hamiltonian.

The JC model can be expanded in numerous ways, of which the most common ones are~\cite{Themis}: {\it multi-modes} where more than one boson mode is considered, {\it multi-level atom} in which more than two electronic levels couple to the electromagnetic field, {\it multi-atom} where $N$ identical two-level atoms couple to the boson mode (the so called Dicke model), or {\it pumped/driven} models. Focusing on the first extension for now (and return to the other below), and labelling each boson mode by an index $i$ and assuming that the atom couples to each bosonic mode with a corresponding coupling strength $g_i$, the multi-mode JC model takes the from
\begin{equation}
    \hat{H}=\frac{\Omega}{2}\hat{\sigma}^z+\sum_i\omega_i \hat{n}_i+ \sum_i g_i\left(\hat{\sigma}^+\hat{a}_i+\hat{\sigma}^-\hat{a}_i^{\dagger}\right).\label{eq:Multi-JC}
\end{equation}
Like for the regular JC model we could turn to an interaction picture with respect to the total excitation number, which is especially convenient in the degenerate case when $\omega_i\equiv\omega,\,\,\,\forall\,\,i$. 

\subsection{Fock state lattices for the resonant multi-mode Jaynes-Cummings models}
From the single mode JC model we first see how the lattice like structure emerges in state space. As long as the coupling $g\neq0$, the Fock states are in general not the eigenstates of the JC model, instead they provide a natural basis for analyzing the dynamics of the system. Consider first the Fock states for the single mode JC model, $\ket{n,e}$ and $\ket{n,g}$. Letting the Hamiltonian in~(\ref{JC-Ham}) act on these states one finds that they, except for the vacuum state, will couple pairwise as
\begin{equation}\label{jctun}
    \ket{n,g} \longleftrightarrow\ket{n-1,e}.
\end{equation}
Thus, the JC Hamiltonian is $2\times2$ block diagonal in the Fock basis. 

It is important to appreciate that the Fock states are coupled by the interaction Hamiltonian~(\ref{jcintham}) alone, while the bare part $\hat H_0$ of the Hamiltonian acts only as onsite energy shifts in the FSL. Throughout this paper we break up the Hamiltonian of the problem as $\hat H=\hat H_0+\hat H_\mathrm{int}$, where the first part only contributes to onsite energy shifts but not any kinematics, which instead is contained within the interacting part $\hat H_\mathrm{int}$. Indeed, the lattice geometry and its tunneling rates are defined via $\hat H_\mathrm{int}$, and we will thereby focus on the lattices deriving from the interaction Hamiltonians without paying any attention on the energy shifts. In some occasions the bare part will only contribute to an overall energy shift, while in others its role is more delicate and in such cases we imagine working in an interaction picture.

\begin{figure}[htb]
    \centering % <-- added
\begin{subfigure}{}
  \begin{tikzpicture}
        \draw (-1,1) node []{$(\textbf{a})$};
        \foreach \rowa in {0, 1} {
        \foreach \rowb in {0} {
        \draw ($2*(\rowa,\rowb)+(1,0)$) -- ($2*(\rowa,\rowb)+(1,2)$)[draw=black,line width=1.5pt];
    }
    }
        \foreach \rowa in {0, 1} {
        \draw ($2*(\rowa,1)+(1,0)$) node [ellipse,fill=red, draw=red,minimum width=7pt,minimum height=7pt]{};
        \draw ($2*(\rowa,0)+(1,0)$) node [ellipse,fill=blue, draw=blue,minimum width=7pt,minimum height=7pt]{};
    }
    
    \draw (1.9,1) node []{$g\sqrt{n-1}$};
    \draw (3.9,1) node []{$g\sqrt{n}$};
    \draw (1,-0.5) node []{$|n-1,g\rangle$};
    \draw (3,-0.5) node []{$|n,g\rangle$};
    \draw (1,2.5) node []{$|n-2,e\rangle$};
    \draw (3,2.5) node []{$|n-1,e\rangle$};
\end{tikzpicture}
  %\label{fig:1}
\end{subfigure}\hfil\hfil % <-- added
\begin{subfigure}{}
\scalebox{1}{
\begin{tikzpicture}
 % For a hexagons the coordinates for the vertices in a sequence:
 %  (-1,0), (-1/2,sqrt{3}/2), (1/2,sqrt{3}/2), (1,0), (1/2,-sqrt{3}/2), (-1/2,-sqrt{3}/2)
 % sqrt{3}/2 = 0.866 (approx)   
    \draw (-4.2,1) node []{$(\textbf{b})$};
    \draw ($(-4,0)$)  --($(-2,0)$)[draw=orange,line width=1.5pt];
    \draw ($(0,0)$)  --($(2,0)$)[draw=orange,line width=1.5pt];
    
    \draw ($(1,1.5)$) node {$|0,n-1,e\rangle$};
    \draw ($(-3,1.5)$) node {$|n-1,0,e\rangle$};
    
    \draw ($(2.4,-0.5)$) node {$|0,n,g\rangle$};
    \draw ($(-4.4,-0.5)$) node {$|n,0,g\rangle$};
    \draw ($(-1.8,-0.5)$) node {$|n-1,1,g\rangle$};
    \draw ($(-0.0,-0.5)$) node {$|1,n-1,g\rangle$};
    \draw ($(-0.8,0.5)$) node {{\huge $\cdots$}};
  
    \foreach \rowa in {-2} {
   \draw ($2*(\rowa,0)$)--($(1,0)+(2*\rowa,0.866)$)[draw=black,line width=1.5pt];
   \draw ($(2,0)+2*(\rowa,0)$)--($(1,0)+(2*\rowa,0.866)$)[draw=black,line width=1.5pt];
   %\draw ($(1,1)+(2*\rowa,0.866)$)--($(1,0)+(2*\rowa,0.866)$)[draw=black,line width=1.5pt];
   \draw ($(1,0)+(2*\rowa,0.866)$) node [ellipse,fill=red, draw=red,minimum width=2pt,minimum height=2pt]{};
    }
    
    \foreach \rowa in {0} {
   \draw ($2*(\rowa,0)$)--($(1,0)+(2*\rowa,0.866)$)[draw=black,line width=1.5pt];
   \draw ($(2,0)+2*(\rowa,0)$)--($(1,0)+(2*\rowa,0.866)$)[draw=black,line width=1.5pt];
   %\draw ($(1,1)+(2*\rowa,0.866)$)--($(1,0)+(2*\rowa,0.866)$)[draw=black,line width=1.5pt];
   \draw ($(1,0)+(2*\rowa,0.866)$) node [ellipse,fill=red, draw=red,minimum width=2pt,minimum height=2pt]{};
    }
    \foreach \rowa in {-2} {
   \draw ($2*(\rowa,0)$) node [ellipse,fill=blue, draw=blue,minimum width=2pt,minimum height=2pt]{};
    }
        \foreach \rowa in {1} {
   \draw ($2*(\rowa,0)$) node [ellipse,fill=blue, draw=blue,minimum width=2pt,minimum height=2pt]{};
    }
    
 \foreach \rowa in {0} {
   \draw ($2*(\rowa,0)$) node [ellipse,fill=blue, draw=blue,minimum width=2pt,minimum height=2pt]{};
    }
          \foreach \rowa in {-1} {
   \draw ($2*(\rowa,0)$) node [ellipse,fill=blue, draw=blue,minimum width=2pt,minimum height=2pt]{};
    }
    
\end{tikzpicture}}
  %\label{fig:2}
\end{subfigure}\hfil % <-- added   
\caption{The FSL for the single-mode JC (a) and two-mode JC (b) models, with red/blue dots marking the atomic $|e/g\rangle$ states. In the former one finds an infinite series of two-site lattices, and in the latter one has a 1D chain. The black lines indicate the non-zero tunnelings of the lattice in the resonant limit, whereas the orange ones (in the two-mode model) are the same but in the large detuned limit when the excited atomic level $\ket{e}$ has been adiabatically eliminated. In both cases the tunneling rates scale such that the lattice is not translationally invariant ($\sqrt{n_i}$ in the resonant case -- black lines, and $\sqrt{n_an_b}$ in the large detuning limit -- orange lines).}
\label{fig:JC-1}
\end{figure}
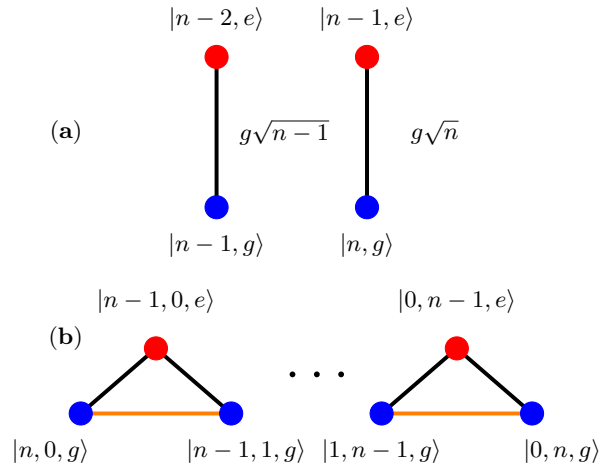

Returning to the JC model, the Hamiltonian decouples into blocks, and the related lattice structure becomes rather trivial as shown in Fig.~\ref{fig:JC-1} (a). We find an infinite set of two-site lattices, alternatively a ladder lattice with the tunneling rates along the legs being zero. One should note that as $\hat{a}^\dagger\ket{n}=\sqrt{n}\ket{n+1}$, the tunneling rate scales with $\sqrt{n}$ for the two-site lattice $n$. This is general for all our models, i.e. the tunnelings are not constant throughout the lattices, and hence the lattices do not possess translational invariance. A consequence is that Bloch's theorem does not apply and we typically find a point spectrum, rather than a band spectrum.

For the two-mode JC model a similar, but still conceptually different, picture arises. The interaction Hamiltonian in Eq.~(\ref{eq:Multi-JC}) reduces to
\begin{equation}
        \hat{H}_\mathrm{int}= \left(g_A\hat{\sigma}^+\hat{a}+g_B\hat{\sigma}^+\hat{b}+h.c\right).\label{eq:2mode-JC}
\end{equation}
 As there are now two boson modes the Fock states will take the form $\ket{n_A,n_B, g}$ and $\ket{n_A,n_B,e}$. According to Eq.~(\ref{eq:2mode-JC}) the states couple as
\begin{equation}
\begin{split}
    ..\ket{n_A,n_B-1,e}\leftrightarrow\ket{n_A,n_B,g}\leftrightarrow\ket{n_A-1,n_B,e}..\, .
    \end{split}
\end{equation}
Hence, for the two-mode JC model, a lattice like structure of two identical copies of 1D chains results, see Fig.\ref{fig:JC-1} (b). As for the single mode JC model, the couplings scale with $\sqrt{n}$. In the two-mode JC model the atom can mediate photon transfer between the two modes~\cite{Themis}, which for the FSL implies moving left or right along the 1D lattice.

%From Fig. \ref{fig:JC-1} it is clear that for any $D$-mode JC model, the lattice like structure that emerges will consist of infinitely many copies of of a similar $D-1$ dimensional lattice like structure, where each copy keeps the number of excitations of spins and boson modes constant. 

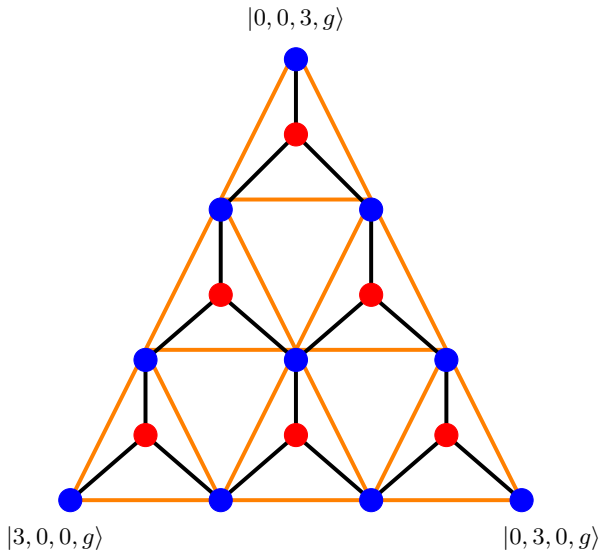
\begin{figure}[htb]
    %\centering % <-- added
 \begin{tikzpicture}
 % For a hexagons the coordinates for the vertices in a sequence:
 %  (-1,0), (-1/2,sqrt{3}/2), (1/2,sqrt{3}/2), (1,0), (1/2,-sqrt{3}/2), (-1/2,-sqrt{3}/2)
 % sqrt{3}/2 = 0.866 (approx)   
      
    \draw ($(0,0)$)  --($(2,0)$)[draw=orange,line width=1.5pt];
    \draw ($(0,0)$)  --($(-2,0)$)[draw=orange,line width=1.5pt];
    \draw ($(4,0)$)  --($(2,0)$)[draw=orange,line width=1.5pt];
    \draw ($(0,0)$)  --($(1,2)$)[draw=orange,line width=1.5pt];
    \draw ($(-2,0)$)  --($(-1,2)$)[draw=orange,line width=1.5pt];
    \draw ($(-1,2)$)  --($(1,2)$)[draw=orange,line width=1.5pt];
    \draw ($(3,2)$)  --($(1,2)$)[draw=orange,line width=1.5pt];
    \draw ($(2,0)$)  --($(1,2)$)[draw=orange,line width=1.5pt];
    \draw ($(2,0)$)  --($(3,2)$)[draw=orange,line width=1.5pt];
    \draw ($(3,2)$)  --($(4,0)$)[draw=orange,line width=1.5pt];
    \draw ($(-1,2)$)  --($(0,0)$)[draw=orange,line width=1.5pt];
    \draw ($(-1,2)$)  --($(0,4)$)[draw=orange,line width=1.5pt];
    \draw ($(1,2)$)  --($(0,4)$)[draw=orange,line width=1.5pt];
    \draw ($(2,4)$)  --($(1,2)$)[draw=orange,line width=1.5pt];
    \draw ($(2,4)$)  --($(3,2)$)[draw=orange,line width=1.5pt];
    \draw ($(2,4)$)  --($(1,6)$)[draw=orange,line width=1.5pt];
    \draw ($(0,4)$)  --($(2,4)$)[draw=orange,line width=1.5pt];
    \draw ($(0,4)$)  --($(1,6)$)[draw=orange,line width=1.5pt];
    \draw ($(4.4,-0.5)$) node {$|0,3,0,g\rangle$};
    \draw ($(-2.2,-0.5)$) node {$|3,0,0,g\rangle$};
    \draw ($(1,6.4)$) node {$|0,0,3,g\rangle$};
    
    \foreach \rowa in {1} {
    \draw ($(-1,4)+(2*\rowa,0.866)$)  --($(-1,5)+(2*\rowa,0.866)$)[draw=black,line width=1.5pt];
    \draw ($(-1,4)+(2*\rowa,0.866)$)  --($(0,3)+(2*\rowa,0.866)$)[draw=black,line width=1.5pt] ;
    \draw ($(-1,4)+(2*\rowa,0.866)$)  --($(-2,3)+(2*\rowa,0.866)$) [draw=black,line width=1.5pt];
    \draw ($(-1,4)+(2*\rowa,0.866)$) node [ellipse,fill=red, draw=red,minimum width=2pt,minimum height=2pt]{};
    }
    \foreach \rowa in {1} {
    \draw ($(-1,5)+(2*\rowa,0.866)$) node [ellipse,fill=blue, draw=blue,minimum width=2pt,minimum height=2pt]{};
    }
    \foreach \rowa in {0, ...,1} {
    \draw ($(0,1)+2*(\rowa,0.866)$) --($(-1,1)+(2*\rowa,0.866)$)[draw=black,line width=1.5pt];
    \draw ($(0,1)+2*(\rowa,0.866)$) --($(1,1)+(2*\rowa,0.866)$)[draw=black,line width=1.5pt];
    \draw ($(0,1)+2*(\rowa,0.866)$) --($(0,3)+(2*\rowa,0.866)$)[draw=black,line width=1.5pt];
    \draw ($(0,1)+2*(\rowa,0.866)$) node [ellipse,fill=red, draw=red,minimum width=2pt,minimum height=2pt]{};
    }
    \foreach \rowa in {0, ...,1} {
    \draw ($(0,3)+(2*\rowa,0.866)$) node [ellipse,fill=blue, draw=blue,minimum width=2pt,minimum height=2pt]{};
    }
    
    \foreach \rowa in {-1, ...,1} {
   \draw ($2*(\rowa,0)$)--($(1,0)+(2*\rowa,0.866)$)[draw=black,line width=1.5pt];
   \draw ($(2,0)+2*(\rowa,0)$)--($(1,0)+(2*\rowa,0.866)$)[draw=black,line width=1.5pt];
   \draw ($(1,1)+(2*\rowa,0.866)$)--($(1,0)+(2*\rowa,0.866)$)[draw=black,line width=1.5pt];
   \draw ($(1,0)+(2*\rowa,0.866)$) node [ellipse,fill=red, draw=red,minimum width=2pt,minimum height=2pt]{};
    }
    
    \foreach \rowa in {-1, ...,1} {
    
   \draw ($(1,1)+(2*\rowa,0.866)$) node [ellipse,fill=blue, draw=blue,minimum width=2pt,minimum height=2pt]{};
    }
    
    \foreach \rowa in {-1, ...,2} {
   \draw ($2*(\rowa,0)$) node [ellipse,fill=blue, draw=blue,minimum width=2pt,minimum height=2pt]{};
    }
\end{tikzpicture}
\caption{Same as for Fig.~\ref{fig:JC-1} but for the three-mode JC model. Here we show it for the $N=3$ excitation sector, which results in a finite lattice comprising 16 sites (increasing $N$ makes the lattice larger). In the resonant case, the FSL forms a hexagonal structure in which each site couples to three neighbours (envisioned by the solid black lines), while in the detuned limit one of the atomic states (red or blue dotes) are adiabatically eliminated and a triangular FSL derives, where every site couples to six neighbouring sites instead (shown by the solid orange lines).}
\label{jc3m}
\end{figure}

Finally we consider the FSL for the three-mode JC model, which was first introduced in Ref.~\cite{FSL1}. For a single mode there is only a single neighbouring state to tunnel to (see Eq.~(\ref{jctun})), while for two modes there are two neighbouring states, and we then build up a 1D chain. For three modes, $a$, $b$, and $c$, we instead have three neighbouring sites, and as such the resulting lattice will be hexagonal as shown in Fig.~\ref{jc3m}. The hexagonal lattice has a sublattice triangular structure seen as the red and blue dots in the figure, which further represent the different atomic states; $|g\rangle$ (blue) and $|e\rangle$ (red). These triangular sublattices will be those surviving in the large detuning limit as shown in the next section. The translationally invariant hexoganl lattice has several interesting features~\cite{jonasbook}, e.g. the presence of Dirac cones at the corners of the Brillouin zone. Furthermore, if tunneling beyond nearest neighbour is considered (Haldane model), the model may host topological features which was analyzed for the FSL in~\cite{jctop}. You could generalize this construction of the FSLs to the $n$-mode JC model which will result in a lattice in an $(n-1)$-dimensional space. 

\subsection{Beyond the rotating wave equation -- the quantum Rabi model}
\begin{figure}[htb]
%\begin{tikzpicture}{R}{6cm}
     \centering
  \begin{tikzpicture}

        \draw ($2*(0,0)$) -- ($2*(-1,0)$)[draw=black,line width=1pt];
        \draw [-stealth]($2*(0,0)$) -- ($2*(-0.7,0)$)[draw=black,line width=1pt];
        \draw [-stealth]($2*(0,0)$) -- ($2*(0.7,0)$)[draw=black,line width=1pt];
        \draw [-stealth]($2*(0,0)$) -- ($2*(-0.5,0.5)$)[draw=black,line width=1pt];
        \draw [-stealth]($2*(0,0)$) -- ($2*(0.5,0.5)$)[draw=black,line width=1pt];
        \draw [-stealth]($2*(0,0)$) -- ($2*(0,0.5)$)[draw=black,line width=1pt];
        \draw ($2*(0,0)$) -- ($2*(0,1)$)[draw=black,line width=1pt];
        \draw ($2*(0,0)$) -- ($2*(-1,1)$)[draw=black,line width=1pt];
        \draw ($2*(0,0)$) -- ($2*(1,1)$)[draw=black,line width=1pt];
        \draw ($2*(0,0)$) -- ($2*(1,0)$)[draw=black,line width=1pt];

        \foreach \rowa in {-1,0, 1} {
        \foreach \rowb in {0, 1} {
        \draw ($2*(\rowa,\rowb)$) node [ellipse,fill=black, draw=black,minimum width=7pt,minimum height=7pt]{};
    }
    }
    \draw (0.3,1.3) node []{$\sigma^{+}$};
    \draw (-2,1.3) node []{$\hat{a}\sigma^{+}$ (I)};
    \draw (2,1.3) node []{$\hat{a}^{\dagger}\sigma^{+}$ (II)};
    \draw (-1,0.3) node []{$\hat{a}$ (III)};
    \draw (1,0.3) node []{$\hat{a}^{\dagger}$ (III)};
    
    \draw (-2,-0.5) node []{$|n-1,g\rangle$};
    \draw (0,-0.5) node []{$|n,g\rangle$};
    \draw (2,-0.5) node []{$|n+1,g\rangle$};
    \draw (-2,2.5) node []{$|n-1,e\rangle$};
    \draw (0,2.5) node []{$|n,e\rangle$};
    \draw (2,2.5) node []{$|n+1,e\rangle$};
\end{tikzpicture}
         \caption{Schematic representation of the zeroth and first order couplings of Fock states. The order of the coupling refers to the number of participating boson creation or annihilation operators in each tunneling process.}
         \label{fig:neighbors}
%\end{tikzpicture}
\end{figure}
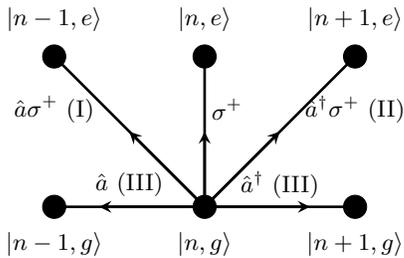
 Returning to the Hamiltonian of Eq.~(\ref{eq:RWA}), we are immediately presented with another lattice model if we keep the counter rotating terms rather than dismiss them. This model is called the quantum Rabi model~\cite{Themis,parity}. More compactly, the interaction Hamiltonian can be expressed as $\hat{H}_{int}=g(\hat{a}^\dagger+\hat{a})\hat{\sigma}_x$, where $\hat\sigma_x$ is the Pauli-$x$ matrix; $\hat\sigma_x|g(e)\rangle=|g(e)\rangle$. This Hamiltonian does not preserve the number of excitations as the JC Hamiltonian of Eq.~(\ref{JC-Ham}). The counter rotating terms break down the continuous symmetry of the JC model to a discrete $\mathbb{Z}_2$-parity symmetry~\cite{parity}. We will discuss the role of symmetries in more general terms below. For the FSL, these terms change the number of excitations by one, i.e. they connect the $2\times2$ blocks of the JC Hamiltonian. In addition, each Fock state now couples to two neighboring Fock states;
\begin{equation}
    \dots\ket{e,-2}\longleftrightarrow\ket{g,n-1}\longleftrightarrow\ket{e,n}\longleftrightarrow\ket{g,n+ 1}\dots\,,
\end{equation}
With two neighbours the FSL becomes a 1D chain (see Fig. \ref{fig:FSL}). The aforementioned $\mathbb{Z}_2$-symmetry implies two decoupled parity sectors, meaning that we actually get two copies of the chain. If we consider the anisotropic quantum Rabi model~\cite{Anrabi2}, we have that the counter rotating terms couple with a different coupling amplitude than the excitation preserving terms, i.e. the interaction Hamiltonian takes the form
\begin{equation}\label{anRabi}
\hat{H}_\mathrm{int}=g_\mathrm{jc}\left(\hat{\sigma}_{+}\hat{a}+\hat{a}^\dagger\hat{\sigma}_{-}\right)+g_\mathrm{ajc}\left(\hat{\sigma}_{-}\hat{a}+\hat{a}^\dagger\hat{\sigma}_{+}\right).
\end{equation}
This does not change the lattice geometry, but an alternation between the tunneling rates of every second site occurs. For a translational invariant system this would mean that the lattice had a bi-partite structure where every unit cell would be comprised of two sites. This is the well known Su-Schrieffer–Heeger (SSH) model~\cite{ssh,jonasbook} which has exponetially localized topologically protected zero energy edge states. Similar edge states are also found in the anisotropic quantum Rabi model, but an obstacle here is that the FSL of the anisotropic quantum Rabi model has only one and not two edges (the boson Fock space has no upper bound). Furthermore, even though the edge state is localized to the edge it is not exponentially localized. However, one can introduce another model supporting a finite FSL, which we do in Sec.~\ref{sec:ssh}, in which the edge states come out more natural and they are exponentially localized.

The above serves also as a demonstration how the different coupling terms manifest as tunneling rates in the FSL. This can be done systematically in order to construct other FLSs. In Fig.~\ref{fig:neighbors} the three different couplings of a Fock state consisting of a single boson mode and a spin half excitation are shown. We label the couplings $(I)-(III)$, and refer to them as first order couplings since they all consist of a single boson creation or annihilation operator, and therefore all scale with $\sqrt{n}$. The last coupling shown in Fig.~\ref{fig:neighbors} only comprises a coupling of the spin mode - and no boson operator - and is therefore a zeroth order. We do not consider the purely spin-coupling any further; higher order couplings typically lead to couplings beyond nearest neighbours, and they may lead to new interesting lattice like structure. The $(I)$ term is the JC coupling and the $(II)$ the counter rotating coupling, which we both have already been discussed. The last couplings, marked by $(III)$,  stem from driving the boson mode. The inclusion of the drive term is discussed for the JC model in the section below and for the Quantum Rabi model in Sec.~\ref{sec:resonant}.

\subsection{Driven Jaynes-Cummings like systems}
As mentioned above, the counter rotating terms lowers the symmetry of the JC model, and the $2\times2$ block structure of the Hamiltonian in the Fock basis is lost. Looking at the JC FSL in Fig.~\ref{fig:JC-1} (a), this implies additional coupling terms such that we find two 1D chains; one for each parity. Naturally, we can break the continuous $U(1)$ number conservation symmetry in different ways. A natural way to break this symmetry is to consider the driven JC model, where we include a field drive
\begin{equation}
    \hat H_\mathrm{fd}=\eta\left(\hat a+\hat a^\dagger\right),
\end{equation}
or an atom drive
\begin{equation}
    \hat H_\mathrm{ad}=\eta\hat\sigma_x.
\end{equation}
For either of these, the excitation number is no longer preserved, while both still support a $\mathbb{Z}_2$-parity symmetry. More precisely, the field drive causes non-zero horizontal tunneling in the FSL of Fig.~\ref{fig:JC-1} (a), and the atom drive results in diagonal tunneling terms. As for the anisotropic quantum Rabi model, which mimicked the SSH model with broken translational invariance, the driven quantum Rabi model realizes a Creutz ladder~\cite{cl} with broken translational invariance, see Fig.~\ref{fig:FSL} (d). The Creutz ladder has attracted attention as it may host dispersionless, i.e. flat, bands. Naturally, such flat bands do not occur in the driven quantum Rabi model since it is no longer translationally invariant. However, remnants of these bands are still present. For field driving, the interaction Hamiltonian can be written as
\begin{equation}
    \hat H_\mathrm{int}^{(\pm)}=\left(g\pm\eta\right)\left(\hat a+\hat a^\dagger\right), 
\end{equation}
which results in two energy branches for which one is flat provided $g=\pm\eta$. This, however, is true in the interaction picture; the bare Hamiltonian $\hat H_0$ would lift this degeneracy as well as causing the Hamiltonian to be bounded from below.

\section{Note on symmetries and lattice dimensions}\label{sec:symmetry}
We have seen a couple of examples in which the structure, and especially the dimensionallity, of the lattice is changed if the symmetry of the system is altered. Here we give some general rules for how this comes about.

Returning to Fig. \ref{fig:neighbors} we may also use it to prove a recipe for how one may construct even more elaborate FSLs. First we notice that in this figure we consider a system of a single boson mode and a two-level atom, but we could expand it to multi-level atoms, several boson modes or a combination of these. If we replace the two-level atom by an $S$-level atom, it  will actually provide a pseudo dimension that will be $2S+1$ sites deep, while including another boson mode adds an extra full dimension (infinitely deep). Generally, the bosonic modes, constituting one continuous degree-of-freedom, each gives rise to one lattice dimension, such that with each additional bosonic mode the dimension of the FSL grows with one. The atom, which provides a discrete degree-of-freedom, causes an extra lattice pseudo dimension, i.e. the lattice is finite in the new dimension. 

A continuous $U(1)$ symmetry reduces the lattice dimension by unity, while a discrete $\mathbb{Z}_n$ symmetry decouple the lattice into sublattices. For example, as the JC model consists of one continuous boson degree-of-freedom and a discrete atomic degree-of-freedom, one might expect a ladder with two legs. However, number conservation reduces the lattice dimension by one, which prohibits tunneling along the legs of the ladder. The quantum Rabi model, on the other hand, does not support particle conservation and we allow tunneling along the legs, while the parity symmetry causes the rung tunnelings to be zero. Note, further, that a continuous symmetry can make the size of the lattice finite, apart from reducing the dimension. Examples of these are given in Sec.~\ref{sec:detuned}. 

Given some Hamiltonian, its FSL typically supports some discrete point symmetries. Take for example the triangular FSL of Fig.~\ref{jc3m}, which clearly has a $2\pi/3$ rotational symmetry around its center site, and three mirror reflections. These lattice symmetries can be traced back to symmetries of the Hamiltonian, e.g. the rotational symmetry translates to the unitary transformation
\begin{equation}
    \hat{a}\rightarrow\hat{b},\,\hat{b}\rightarrow\hat{c},\quad \text{and}\quad\hat{c}\rightarrow\hat{a}
\end{equation}

\begin{figure}[htb]
    %\centering % <-- added
\begin{subfigure}{}
  \scalebox{0.8}{
  \begin{tikzpicture}
        \foreach \rowa in {0, 1, ...,3} {
        \foreach \rowb in {0} {
        \draw ($2*(\rowa,\rowb)$) -- ($2*(\rowa,\rowb)+(0,2)$)[draw=black,line width=1.5pt];
    }
    }
        \foreach \rowa in {0, 1, ...,3} {
        \foreach \rowb in {0, 1} {
        \draw ($2*(\rowa,\rowb)$) node [ellipse,fill=black, draw=black,minimum width=7pt,minimum height=7pt]{};
    }
    }
    \draw (-1,1) node []{$(a)$};
    \draw (0,-0.5) node []{$|n-1,g\rangle$};
    \draw (2,-0.5) node []{$|n,g\rangle$};
    \draw (4,-0.5) node []{$|n+1,g\rangle$};
    \draw (6,-0.5) node []{$|n-2,g\rangle$};
    \draw (0,2.5) node []{$|n-2,e\rangle$};
    \draw (2,2.5) node []{$|n-1,e\rangle$};
    \draw (4,2.5) node []{$|n,e\rangle$};
    \draw (6,2.5) node []{$|n-1,e\rangle$};
    %\draw (3,-1) node []%{The Jaynes Cummings model};
\end{tikzpicture}}
  \label{fig:JCmodel}
\end{subfigure}\hfil % <-- added
\begin{subfigure}{}
 \scalebox{0.8}{\begin{tikzpicture}
 \draw (-1,1) node []{$(b)$};
 
        \foreach \rowa in {0, 1, ...,2} {
        \foreach \rowb in {0} {
        \draw ($(2,0)+2*(\rowa,\rowb)$) -- ($2*(\rowa,\rowb)+(0,2)$)[draw=black,line width=1.5pt];
        \draw ($(0,0)+2*(\rowa,\rowb)$) -- ($(2,0)+2*(\rowa,\rowb)+(0,2)$)[draw=red,line width=1.5pt];
    }
    }
        \draw ($2*(3,1)$) -- ($(6.5,1.5)$)[draw=black,line width=1.5pt];
        \draw ($2*(3,0)$) -- ($(6.5,0.5)$)[draw=red,line width=1.5pt];
        \draw ($2*(0,0)$) -- ($(-0.5,0.5)$)[draw=black,line width=1.5pt];
        \draw ($2*(0,1)$) -- ($(-0.5,1.5)$)[draw=red,line width=1.5pt];
        \foreach \rowa in {0, 1, ...,3} {
        \foreach \rowb in {0, 1} {
        \draw ($2*(\rowa,\rowb)$) node [ellipse,fill=black, draw=black,minimum width=7pt,minimum height=7pt]{};
    }
    }
    \draw (0,-0.5) node []{$|n-1,g\rangle$};
    \draw (2,-0.5) node []{$|n,g\rangle$};
    \draw (4,-0.5) node []{$|n+1,g\rangle$};
    \draw (6,-0.5) node []{$|n+2,g\rangle$};
    \draw (0,2.5) node []{$|n-1,e\rangle$};
    \draw (2,2.5) node []{$|n,e\rangle$};
    \draw (4,2.5) node []{$|n+1,e\rangle$};
    \draw (6,2.5) node []{$|n+2,e\rangle$};
\end{tikzpicture}
  }
  \label{fig:quantumrabi}
\end{subfigure}

\medskip
\begin{subfigure}{}
  \scalebox{0.8}{\begin{tikzpicture}
    \foreach \rowa in {0, 1, ...,2} {
        \foreach \rowb in {0, 1} {
        \draw ($2*(\rowa,\rowb)$) -- ($2*(\rowa,\rowb)+(2,0)$)[draw=blue,line width=1.5pt];
    }
    }
        \foreach \rowa in {0, 1, ...,3} {
        \foreach \rowb in {0} {
        \draw ($2*(\rowa,\rowb)$) -- ($2*(\rowa,\rowb)+(0,2)$)[draw=black,line width=1.5pt];
    }
    }
        \foreach \rowa in {0, 1, ...,3} {
        \foreach \rowb in {0, 1} {
        \draw ($2*(\rowa,\rowb)$) node [ellipse,fill=black, draw=black,minimum width=7pt,minimum height=7pt]{};
    }
    }
    
    \draw (-1,1) node []{$(c)$};
    
    \draw (0,-0.5) node []{$|n-1,g\rangle$};
    \draw (2,-0.5) node []{$|n,g\rangle$};
    \draw (4,-0.5) node []{$|n+1,g\rangle$};
    \draw (6,-0.5) node []{$|n+2,g\rangle$};
    \draw (0,2.5) node []{$|n-2,e\rangle$};
    \draw (2,2.5) node []{$|n-1,e\rangle$};
    \draw (4,2.5) node []{$|n,e\rangle$};
    \draw (6,2.5) node []{$|n+1,e\rangle$};
\end{tikzpicture}
  }
  \label{fig:jcmodeldrive}
\end{subfigure}\hfil % <-- added
\begin{subfigure}{}
  \scalebox{0.8}{\begin{tikzpicture}
    \draw (-1,1) node []{$(d)$};
    \foreach \rowa in {0, 1, ...,2} {
        \foreach \rowb in {0, 1} {
        \draw ($2*(\rowa,\rowb)$) -- ($2*(\rowa,\rowb)+(2,0)$)[draw=blue,line width=1.5pt];
    }
    }
        \foreach \rowa in {0, 1, ...,2} {
        \foreach \rowb in {0} {
        \draw ($(2,0)+2*(\rowa,\rowb)$) -- ($2*(\rowa,\rowb)+(0,2)$)[draw=black,line width=1.5pt];
        \draw ($(0,0)+2*(\rowa,\rowb)$) -- ($(2,0)+2*(\rowa,\rowb)+(0,2)$)[draw=red,line width=1.5pt];
    }
    }
        \draw ($2*(3,1)$) -- ($(6.5,1.5)$)[draw=black,line width=1.5pt];
        \draw ($2*(3,0)$) -- ($(6.5,0.5)$)[draw=red,line width=1.5pt];
        \draw ($2*(0,0)$) -- ($(-0.5,0.5)$)[draw=black,line width=1.5pt];
        \draw ($2*(0,1)$) -- ($(-0.5,1.5)$)[draw=red,line width=1.5pt];
        \foreach \rowa in {0, 1, ...,3} {
        \foreach \rowb in {0, 1} {
        \draw ($2*(\rowa,\rowb)$) node [ellipse,fill=black, draw=black,minimum width=7pt,minimum height=7pt]{};
    }
  
    }
    \draw (0,-0.5) node []{$|n-1,g\rangle$};
    \draw (2,-0.5) node []{$|n,g\rangle$};
    \draw (4,-0.5) node []{$|n+1,g\rangle$};
    \draw (6,-0.5) node []{$|n+2,g\rangle$};
    \draw (0,2.5) node []{$|n-1,e\rangle$};
    \draw (2,2.5) node []{$|n,e\rangle$};
    \draw (4,2.5) node []{$|n+1,e\rangle$};
    \draw (6,2.5) node []{$|n+2,e\rangle$};
\end{tikzpicture}
  }
  \label{fig:quantumrabidrive}
\end{subfigure}\hfil % <-- added
\caption{Summary of of FSLs for models consisting of only a single boson and two-level atomic degree-of-freedom. In (a) the lattice of the JC model~(\ref{jcintham}) is shown. Here, the conservation of $\hat N$ reduces the dimension of the system such that an infinite set of two-site lattices emerges. The second plot (b) gives the FSL for the quantum Rabi model~(\ref{anRabi}), with red lines marking the tunneling due to the counter rotating terms and the black lines the JC interaction terms. Note that this is actually two copies of 1D chains, even though they are drawn in this zig-zag way. The FSL of the driven JC model is shown in (c). The tunneling stemming from the driving are marked by the blue lines, i.e. by removing them we return to JC lattice (a). Finally in (d) we display the FSL of the driven quantum Rabi model (again, the lattice in (b) is regained by omitting the blue lines). This is a type of Creutz lattice with known interesting properties (see main text). A drive of the spin instead of the boson would result in vertical blue lines rather than horizontal ones. }
\label{fig:FSL}
\end{figure}
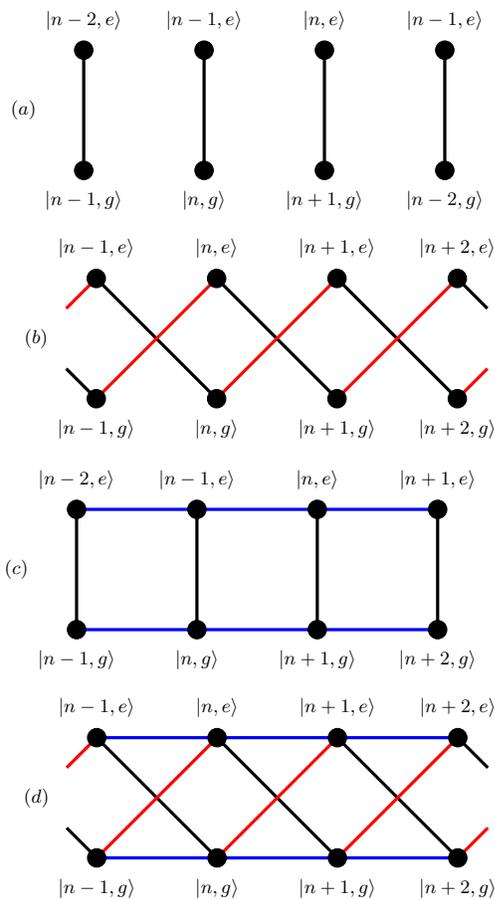

The lattice geometry, e.g. number of neighbours, is determined by the specific form of the interaction Hamiltonian. When expressed in terms of raising/lowering operators (e.g. $\hat a$ and $\hat\sigma^-$), the number of terms in $\hat H_\mathrm{int}$ gives the maximum number of possible neighbours. In the quantum Rabi model there are four terms, but only two cause coupling to other states since $\hat\sigma^+\ket{e}=\hat\sigma^-\ket{g}=0$. The interaction terms may represent tunneling either between nearest neighbours in the FSL or beyond, like for the Creutz ladder as shown in Fig.~\ref{fig:FSL} (d). 

%\begin{itemize}
    %\item \textit{Geometry}: The shape of the base tilling - is the size of the loops in 2D and higher lattices 3,4 or six (or even something else). 
    %\item \textit{Symmetry}: Reduces the dimensions of the lattice
    %\item \textit{High dimensional lattices}: The structure that emerges in 4 or higher dimensions with the given geometry and symmetries.
%\end{itemize}
 
%We are in this article mostly interested in the later, but an example on how higher spins can give rise to an extra dimension is shown in fig \ref{fig:JC-1D}.d. Here a spin degree-of-freedom with three internal states are used to realize a 2D lattice structure. 
%This lattice structure is called a \textit{Lieb}-lattice\cite{Lieb}
%\newpage

\section{The large detuning limit -- quadratic bosnic models}\label{sec:detuned}
\subsection{Effective bosonic models}
Quadratic spin-boson models are intrinsically non-linear due to the finite spin Hilbert spaces. However, there is a natural regime in which the models become approximately linear, namely in the large detuning limit. Here we assume $\left|\Delta\right|\gg g\sqrt{n}$. The fast timescale is then set by $\Delta$, and we face the situation where the atomic degree-of-freedom follows that of the boson. Hence, it is enough to only consider the evolution of the bosons. The effective Hamiltonian that dictates the bosonic dynamics can be derived by adiabatic elimination~\cite{adel}. Here one first writes down the Heisenberg equations for both the boson and spin operators, which for the JC model read
\begin{equation}
\begin{array}{lll}
    \partial_t \hat a & = &  -i\left[\hat a,\hat{H}_\mathrm{JC}\right]=-i g\hat{\sigma}^{-},\\ \\

    \partial_t \hat{\sigma}^{-} & = &-i\Delta \hat{\sigma}^{-}+ig\hat{a}\hat{\sigma}_{z},\\ \\

    \partial_t\hat{\sigma}_z & = & 2ig\left(\hat{a}^{\dagger}\hat{\sigma}^-\right).
    \end{array}
\end{equation}
As the fast timescale is set by $\Delta^{-1}$, the spin operators can be replaced by their steady state solutions, i.e. $\hat{\sigma}^{-}=\frac{g}{\Delta}\hat{a}\hat{\sigma}_{z}$, and after inserting this into the first equation above, the effective dynamical equation for $\hat{a}$ becomes
\begin{equation}
    \partial_t \hat a = -\frac{g^2}{\Delta} \hat a \hat{\sigma}^{z}\equiv -i\left[\hat{a},\hat{H}_\mathrm{eff}\right],
\end{equation}
where in the second step we define the effective bosonic Hamiltonian $\hat{H}_\mathrm{eff}$. Solving this operator equation gives
\begin{equation}
    \hat{H}_\mathrm{eff}=\frac{2g^2}{\Delta}\hat{n}\hat{\sigma}^z.
\end{equation}
In the case of more than one boson modes, the adiabatic elimination is straightforward to generalize. With the summation over an index $i$ on the bosonic operator, the effective Hamiltonian stays the same such that 
\begin{equation}
\begin{array}{lll}
    \partial_t \hat a_i & = & -i g_i\hat{\sigma}^{-},\\ \\
    \partial_t \hat{\sigma}^{-} & =  &  \displaystyle{-i\Delta \hat{\sigma}^{-}+i\sum_i g_i\hat{a}_i\hat{\sigma}_{z}},\\ \\
    \partial_t\hat{\sigma}_z & = & \displaystyle{2i\sum_i g_i\hat{a}_i^{\dagger}\hat{\sigma}^-}.
    \end{array}
\end{equation}
Hence, with $\partial_t \hat{\sigma}^{-}= 0$ we obtain $\hat{\sigma}^{-}=\sum_i \frac{g_i\hat{a}_i\hat{\sigma}_{z}}{\Delta}$ such that 
\begin{equation}
    \hat{H}_\mathrm{eff}=\sum_{i\neq j}\frac{2g_ig_j}{\Delta}\hat{a}^\dagger_i\hat{a}_j\hat{\sigma}^z.\label{eq:eff-Ham}
\end{equation}
From a perturbative perspective, the effective Hamiltonians describe the virtual processes of absorbing and emitting a photon -- the Stark shift. In the multi-mode case, such two-photon processes act as beam-splitters among the involved modes such that bosons will tunnel from one mode to another. This manifests as non-vanishing tunneling rates in the corresponding FSLs.

\subsection{Fock state lattices for multi-mode JC models in the large detuning limit}
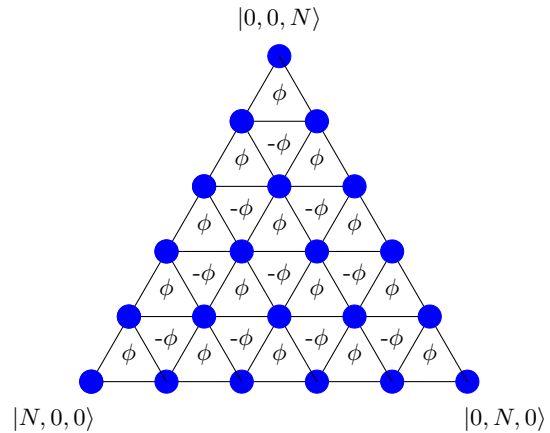
\begin{figure}[htb]
    \centering % <-- added
  \begin{tikzpicture}
    \foreach \row in {0, 1, ...,5} {
        \draw ($\row*(0.5, {0.5*sqrt(3)})$) -- ($(\rows,0)+\row*(-0.5, {0.5*sqrt(3)})$);
        \draw ($\row*(0.5, {0.5*sqrt(3)})$) node[ellipse,fill=blue, draw=blue,minimum width=.5pt,minimum height=0.5pt]{};
        \draw ($\row*(1, 0)$) -- ($(\rows/2,{\rows/2*sqrt(3)})+\row*(0.5,{-0.5*sqrt(3)})$);
        \draw ($\row*(1, 0)$) node[ellipse,fill=blue, draw=blue,minimum width=.5pt,minimum height=0.5pt]{};
        \draw ($\row*(1, 0)$) -- ($(0,0)+\row*(0.5,{0.5*sqrt(3)})$);
        
    }
    \foreach \row in {1, ...,5} {
        \draw ($\row*(0.5, {0.5*sqrt(3)})-(-0,0.5)$) node {$\phi$};
        \draw ($(-0.5, {0.5*sqrt(3)})+\row*(1, 0)$) node[ellipse,fill=blue, draw=blue,minimum width=.5pt,minimum height=0.5pt]{};
    }
    \foreach \row in {1, ...,4} {
        \draw ($\row*(0.5, {0.5*sqrt(3)})-(-0.5,0.3)$) node {-$\phi$};
        \draw ($\row*(0.5, {0.5*sqrt(3)})-(-1,0.5)$) node {$\phi$};
        \draw ($(0, {sqrt(3)})+\row*(1, 0)$) node[ellipse,fill=blue, draw=blue,minimum width=.5pt,minimum height=0.5pt]{};
    }
    
    \foreach \row in {1, ...,3} {
        \draw ($\row*(0.5, {0.5*sqrt(3)})-(-1.5,0.3)$) node {-$\phi$};
        \draw ($\row*(0.5, {0.5*sqrt(3)})-(-2,0.5)$) node {$\phi$};
        \draw ($(0.5, {3*0.5*sqrt(3)})+\row*(1, 0)$) node[ellipse,fill=blue, draw=blue,minimum width=.5pt,minimum height=0.5pt]{};
    }
    
     \foreach \row in {1, ...,2} {
        \draw ($\row*(0.5, {0.5*sqrt(3)})-(-2.5,0.3)$) node {-$\phi$};
        \draw ($\row*(0.5, {0.5*sqrt(3)})-(-3,0.5)$) node {$\phi$};
        \draw ($(1, {4*0.5*sqrt(3)})+\row*(1, 0)$) node[ellipse,fill=blue, draw=blue,minimum width=.5pt,minimum height=0.5pt]{};
    }
    
    \foreach \row in {1} {
        \draw ($\row*(0.5, {0.5*sqrt(3)})-(-3.5,0.3)$) node {-$\phi$};
        \draw ($\row*(0.5, {0.5*sqrt(3)})-(-4,0.5)$) node {$\phi$};
    }

    \draw (-0.5,-0.5) node {$|N,0,0\rangle$};
    \draw (5.5,-0.5) node {$|0,N,0\rangle$};
    \draw ($5*({0.5},{0.5*sqrt(3)})+(0,0.5)$) node {$|0,0,N\rangle$};
\end{tikzpicture}
\hfil % <-- added
\caption{The triangular FSL deriving from the three-mode boson model~(\ref{eq:phase}). A synthetic magnetic flux $\pm\phi$ penetrates each plaquette in a staggered manner. The corner states are represented by the Fock states with two modes in vacuum. Along the straight lines, two diagonal and one horizontal, the boson number is fixed in one of the modes, e.g. for the horizontal lines the third $c$-mode keeps a fixed particle number. These observations holds for higher dimensional lattices as well -- in the four-mode model one finds a tetrahedral lattice.  }
\label{fig:flux}
\end{figure}

With the effective Hamiltonian in Eq.~(\ref{eq:eff-Ham}) we set out to study their properties. 
Noting that in the detuned limit no population is transferred between the atomic states, the spin part of the Fock state becomes trivial and we can leave it out for now. Generally, the interaction Hamiltonian for the detuned multi-mode case can be written on the form $\hat H_\mathrm{int}=\sum_{i\neq j}\tau_{ij}\hat a_i^\dagger\hat a_j$. This describes non-interacting bosons living on a lattice defined by the subscripts $i$ and $j$, e.g. for two and three modes we get a bosonic ``dimer'' or ``trimer'' respectively. We, however, are interested in their respective FSLs. {\it A priori}, the tunneling strengths $\tau_{ij}$ depend on the system details. For degenerate modes and appropriate polarizations it is natural to assume that they all have the same amplitudes, i.e. $\tau_{ij}=\tau e^{i\phi_{ij}}$. The phases $\phi_{ij}$ can, in principle, be experimentally controlled~\cite{jctop}. We include these phases since we will see that their presence leads to a plethora of interesting phenomena. Taking this into account, we reach the final Hamiltonian
\begin{equation}
    \hat{H}_\mathrm{int}=\tau\sum_{i\neq j}e^{i\phi_{ij}}\hat{a}^{\dagger}_i\hat{a}_j.\label{eq:phase}
\end{equation}
Note that hermiticity implies that $\phi_{ij}=-\phi_{ji}$, and since the tunneling rate only sets the time scale we pick $\tau=1$ throughout. For two modes (label them with $a$ and $b$) this Hamiltonian reduces to $\hat{H}_\mathrm{int}= \left( e^{i\phi}\hat{a}^{\dagger}\hat{b}+e^{-i\phi}\hat{b}^{\dagger}\hat{a}\right)$ in which the phase can be trivially omitted by a gauge transformation. However, for three or more modes this will not be true in general ~\cite{gauge}. The phases will result in complex tunneling rates in the FSLs, and according to the idea of the Peierls substitution these phases mimic a synthetic magnetic flux penetrating the lattice perpendicularly~\cite{jonasbook}. If a loop in the lattice is formed, the (gauge invariant) magnetic flux $\phi$ through the loop equals the sum of the phases of the corresponding tunneling rates. Typically we are interested in the smallest such loop, a lattice plaquette, and its corresponding flux, see Fig. \ref{fig:flux}. Consider, for example, three boson modes $a$, $b$ and $c$, from Eq.~(\ref{eq:phase}) we have the three phases $\phi_{ab}$, $\phi_{bc}$, and $\phi_{ca}$. Each of these can be changed via a gauge transforation, but the sum $\phi=\phi_{ab}+\phi_{bc}+\phi_{ca}$ remains constant under such transformations. The corresponding FSL is a triangular lattice with a staggered magnetic flux, i.e. alternating $\pm\phi$ fluxes through each triangular plaquette, see Fig. \ref{fig:flux}. If we make the choice that for any $i<j$ then $\phi_{ij}=\phi$, then for all higher dimensional multi-mode FSLs the same pattern will emerge, i.e. any 2D plane/cut of the FSL will display a triangular lattice with a staggered $\pm\phi$ field. We may note in passing, that it is possible to also have a homogeneous field in the FSL, but it turns out that this is only possible for non-quadratic boson models, i.e. interacting many-body systems.

As the Hamiltonian in Eq.~(\ref{eq:phase}) preserves the number of particles, there is a corresponding $U(1)$-symmetry. This lowers the lattice dimension by one, e.g. the three-mode model generates a 2D FSL and so on. Generally for any number of modes,  the unit cell (in a strict sense we do not have unit cells since the lattice is not translationally invariant, but here we think only of the lattice geometry) has the form of a $d$-simplex, where $d$ indicates the number of dimensions, e.g in 2D it is a triangle, in 3D it is a tetrahedron and so on. Furthermore, for a given particle number $N$, the lattice is finite, with the corners represented by the Fock states with all $N$ bosons occupying a single mode and the remaining ones in vacuum. For example, for three modes with $N$ bosons, the number of sites of the FSL becomes $\mathcal{S}=(N+1)(N+2)/2$. Finally, we note that the introduction of a synthetic field breaks time-reversal symmetry. However, for $\phi=\pi/2$ the spectrum becomes symmetric around $E=0$ which is a manifestation of a chiral symmetry given by the unitary $\hat U_\mathrm{C}=\hat K\exp\left(i\pi\hat n_c\right)$, where $\hat K$ stands for complex conjugation, which anti-commutes with the Hamiltonian. 

\subsection{Closer look at the three-mode model -- a fractal spectrum}
In this section we limit ourselves to the three-mode model with a varying flux. Later we comment also on higher dimensional models. In particular, we are interested in the energy spectrum, for the Hamiltonian given in Eq.~(\ref{eq:phase}), as a function of the flux $\phi$. The translational invariant case of a 2D lattice exposed to a constant perpendicular magnetic field results in a fractal spectrum known as the Hofstadter butterfly~\cite{hof}. It is relevant to analyze similar situations for the FSLs. For this objective we use that the Hamitlonian is quadratic and can be written as $\hat H_\mathrm{int}=\hat{\bf a}^\dagger h\hat{\bf a}$, where 
\begin{equation}
    h=\left[
    \begin{array}{ccc}
        0 & e^{i\phi} & e^{i\phi}\\ 
      e^{-i\phi} & 0 & e^{i\phi}\\ 

     e^{-i\phi} & e^{-i\phi} & 0 
  \end{array}\right]
  \label{eq:matrix}
\end{equation}
and $\hat{\bf a}^\dagger=\left(\hat a^\dagger,\hat b^\dagger,\hat c^\dagger\right)$. If we diagonalize $h$, i.e. $d=UhU^{-1}$, the full Hamiltonian is diagonalized as $\hat H_\mathrm{int}=\gamma_0\hat\alpha_0^\dagger\hat\alpha_0+\gamma_1\hat\alpha_1^\dagger\hat\alpha_1+\gamma_2\hat\alpha_2^\dagger\hat\alpha_2$, with $\hat\alpha_k$ ($k=0,\,1,\,2$) the transformed boson operators, and $\gamma_k$ the eigenvalues of $h$. Thus, the spectrum for a single boson is simply the eigenvalues of $h$, while the full spectrum, given $N$ bosons, can be found combinatorically from these single boson energies $\gamma_k$. %For a general number of modes, the matrix $h$ for the Hamiltonian in Eq.~(\ref{eq:phase}) reads
%\begin{equation}
%    h=\hat{\psi}_n^{\dagger}\left(
%    \begin{array}{cccccc}
%        0& & & & &\\ 
%     &0 & & &  \text{\Large{$e^{i\phi}$}} & \\ 
%     & & \ddots & & & \\
%     &  \text{\Large{$e^{-i\phi}$}} & & & \ddots& \\ 
%     & & & & & 0\\ 
%  \end{array}\right)\hat{\psi}_n.
%  \label{eq:matrix}
%\end{equation}
We may note that the structure of $h$ in Eq.~(\ref{eq:matrix}) remains for higher dimensions. Numerically its eigenvalues can be found for large mode numbers, but analytical solutions are only found for the three- and four-mode cases. For the three-mode model, finding the single boson energies reduces to solving the characteristic polynomial $\gamma^3 - 3\gamma-2\cos\phi=0$, which leads to the three energies 
\begin{equation}
    \gamma_k  = 2\cos\left(\frac{\phi-2\cdot\pi\cdot k}{3}\right), \qquad k\in\left\{0,1,2\right\}
    \label{eq:solv}
\end{equation}
Hence, the full spectrum for a general boson number is written as
\begin{equation}
    E_{\bf n} = n_0\gamma_0+n_1\gamma_1+n_2\gamma_2,
\end{equation}
where ${\bf n}=(n_0,n_1,n_2)$ subject to the constraint $n_0 +n_1+n_2=N$. To gain further insight of the spectral structure we consider the energy differences between any two energy levels
\begin{equation}\label{ediff}
    \delta E_{\bf n}=E_{\bf n}-E_{\tilde{\bf n}} = \left(\gamma_1-\gamma_0\right)\!\left(n_1-\tilde{n}_1\right)+\left(\gamma_2-\gamma_0\right)\!\left(n_2-\tilde{n}_2\right)  
\end{equation}
where we have used that we may rewrite $E_{\bf n} = N\gamma_0+ (\gamma_1-\gamma_0)n_1+(\gamma_2-\gamma_0)n_2,$. In three dimensions there exist certain fluxes $\phi$ for which the spectrum becomes equidistant or quasi equidistant, i.e. there then exists a few energies characterizing the involved time-scales. This is opposed to some general $\phi$ which reproduces a whole range of different time-scales. 

\begin{figure}[ht]
  \centering
  % include first image
  \includegraphics[width=8.4cm]{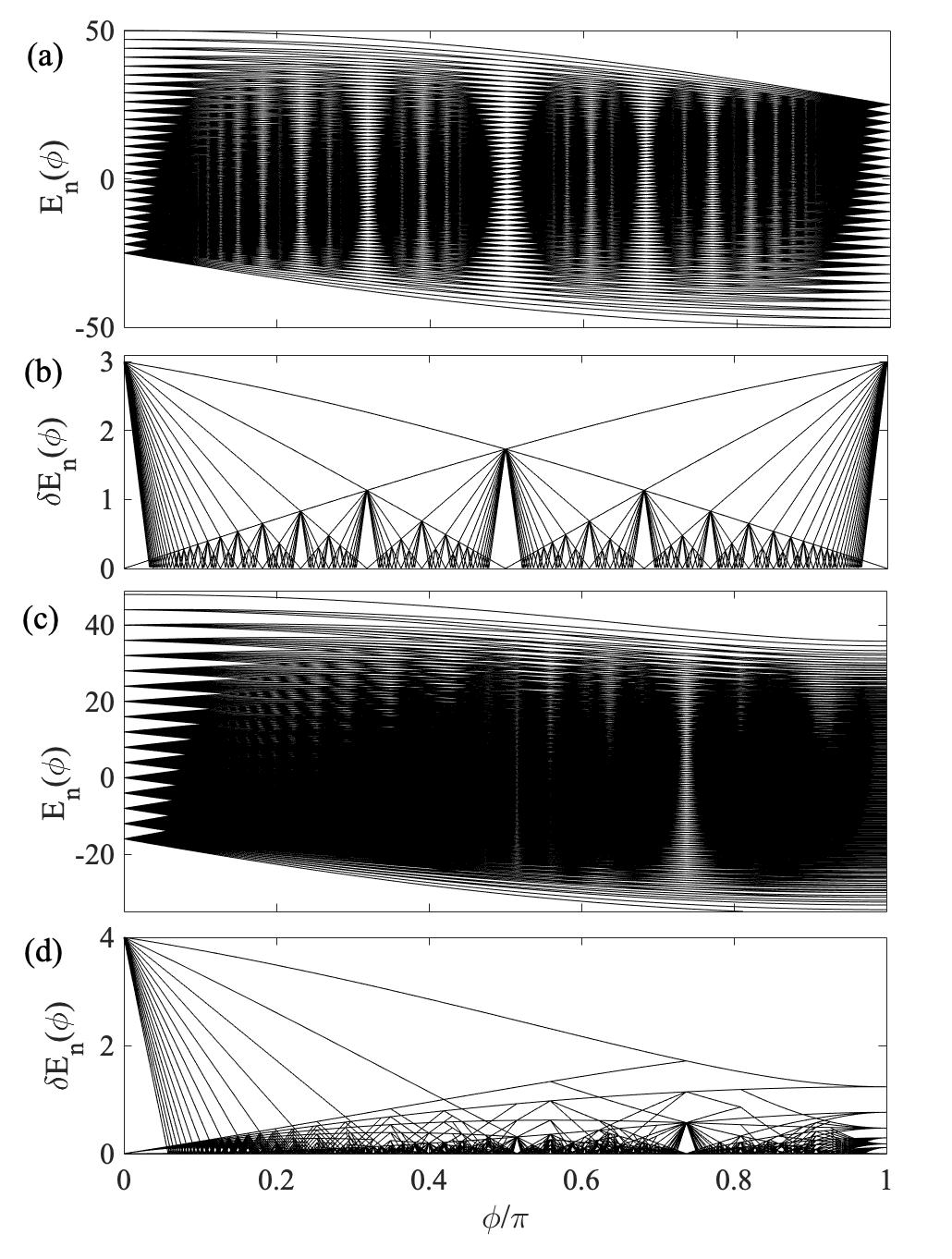}  
\caption{Upper two plots, (a) and (b),  show the spectra and the distances between nearby energies as a function of the flux $\phi$, for the three-mode model with $N=25$ bosons, while the lower two plots, (c) and (d), give the same for the four-mode case with $N=16$. The fractal structure of the three-mode case is evident, while the four-mode spectrum does not display the same fractal properties.  }
\label{fig:fig}
\end{figure}

From the expression~(\ref{ediff}) we see that the energy spectrum becomes quasi equidistant whenever
\begin{equation}
    \frac{\gamma_{0}-\gamma_{1}}{\gamma_{1}-\gamma_{2}}= q, \quad q\in  \mathbb{Q}. 
    \label{eq:fract}
\end{equation}
Writing $q=\frac{j}{j'}$ with $j,j'\in\mathbb{N}$, we find that given the solutions for $\gamma_k$ in~\eqref{eq:solv}, that Eq.~(\ref{eq:fract}) can be written as
\begin{equation}
%\begin{split}
    j'\!\left[\!\sqrt{3}\cos\!\left(\frac{\phi}{3}\right)\!-\!\sin\!\left(\frac{\phi}{3}\right)\!\right]= j\!\left[\!\sqrt{3}\cos\!\left(\frac{\phi}{3}\right)\!+\!\sin\!\left(\frac{\phi}{3}\right)\!\right]\!.
    %\end{split}
    \label{eq:mm}
\end{equation}
As all three single boson energies are continuous functions of $\phi$, so are $(\gamma_{0}-\gamma_{1})$ and $(\gamma_{1}-\gamma_{2})$, and furthermore, except for the singularity when $\gamma_{1}=\gamma_{2}$, which happens at $\phi=0$, also the fraction~(\ref{eq:fract}) is a continuous function of $\phi$. Disregarding this singularity, as $N\rightarrow \infty$ there will be infinitely many $\phi_j$'s that fulfill the condition above, see Fig.~\ref{fig:fig} (a) and (b). The spectrum at these ``quasi equidistant points'' becomes highly degenerate, but it is only perfectly equidistant for one of them, $\phi_1=\pi/2$. This is the reason why for the other of these points we call the spectrum quasi equidistant, since here there exists more than a single energy difference. For a given $\phi_j$, the energy differences between the nearby energies can be ordered $\Delta_j$, $2\Delta_j$, $3\Delta_j$, ..., $j\Delta_j$. Note that the basic energy difference $\Delta_j$ is different for the various equidistant points. Solving~(\ref{eq:fract}) for $\phi_j$ leads to the following condition
\begin{equation}\label{eq:fract2}
    \tan\left(\frac{\phi_j}{3}\right) = \frac{\sqrt{3}}{2j+1}, \quad\quad j\in\mathbb{N}.
\end{equation}
which we have reached using~\eqref{eq:mm}, and introducing a new pair $n,n'\in\mathbb{N}$ where $j+j'=n'$ and $j'-j=n$. As this holds for any $q$ we may choose $n=1$, which implies that $n'=2j+1$ and leads to the expression above. 
 %The first, for example, reads
%\begin{equation}
    %\Delta =\left|\min\left\{2\left[\cos\left(\frac{\phi-2\pi k}{3}\right)-\cos\left(\frac{\phi-2\pi l}{3}\right)\right]\right\}\right|, 
%\end{equation}
%where $ k\neq l \in\left\{0,1,2\right\}$.

The equidistant points given by Eq.~(\ref{eq:fract2}) occur denser and denser the larger values $j$. This manifests as a fractal structure of the spectrum, see Fig.~\ref{fig:fig} showing both the spectrum (a) and its nearby energy differences $\delta E_{\bf n}$ (b). The results of the energy differences demonstrate the concept of the equidistant points. As we will demonstrate in the next section, picking $\phi$ according to one of these points will result in novel dynamics.

\subsection{The four-mode and higher-mode models}
To fractal structure of the spectrum is rooted in the existence of special fluxes fulfilling the condition~(\ref{eq:fract}). The $3\times3$ matrix $h$ gives three energies, and hence two energy difference, and the single condition deriving from comparing these two. For the four-mode model~(\ref{eq:phase}) we find four eigenvalues $\gamma_k$ (with $k=1,2,3,4$) of the corresponding matrix $h$. Following the same arguing that led to the condition~(\ref{eq:fract}), for the four-mode model we are left we three conditions 
\begin{equation}
    \frac{\gamma_3-\gamma_0}{\gamma_2-\gamma_0}=q_1,\qquad \frac{\gamma_3-\gamma_0}{\gamma_1-\gamma_0}=q_2,\qquad 
    \frac{\gamma_2-\gamma_0}{\gamma_1-\gamma_0}=q_3,
\end{equation}
where $q_1$, $q_2$, and $q_3$ are all rational numbers. We can generalize this to a $K$-mode model in which the number of conditions becomes $(K-1)(K-2)/2$. It should be clear that it becomes harder and harder, the larger $K$ is, to find fluxes $\phi_j$ fulfilling all conditions simultaneously. As long as the conditions cannot be met for single fluxes the spectrum cannot be fractal as for the three-mode model. This we demonstrate by displaying the spectrum and energy difference $\delta E_{\bf n}$ for the four-mode model in Fig.~\ref{eq:fract} (c) and (d). There is, however, possible to find fractal spectra also for higher mode models (four and six modes) as discussed next.

\subsubsection{Ring models}
\begin{figure}[ht]
  \centering
  % include first image
  \includegraphics[width=8.4cm]{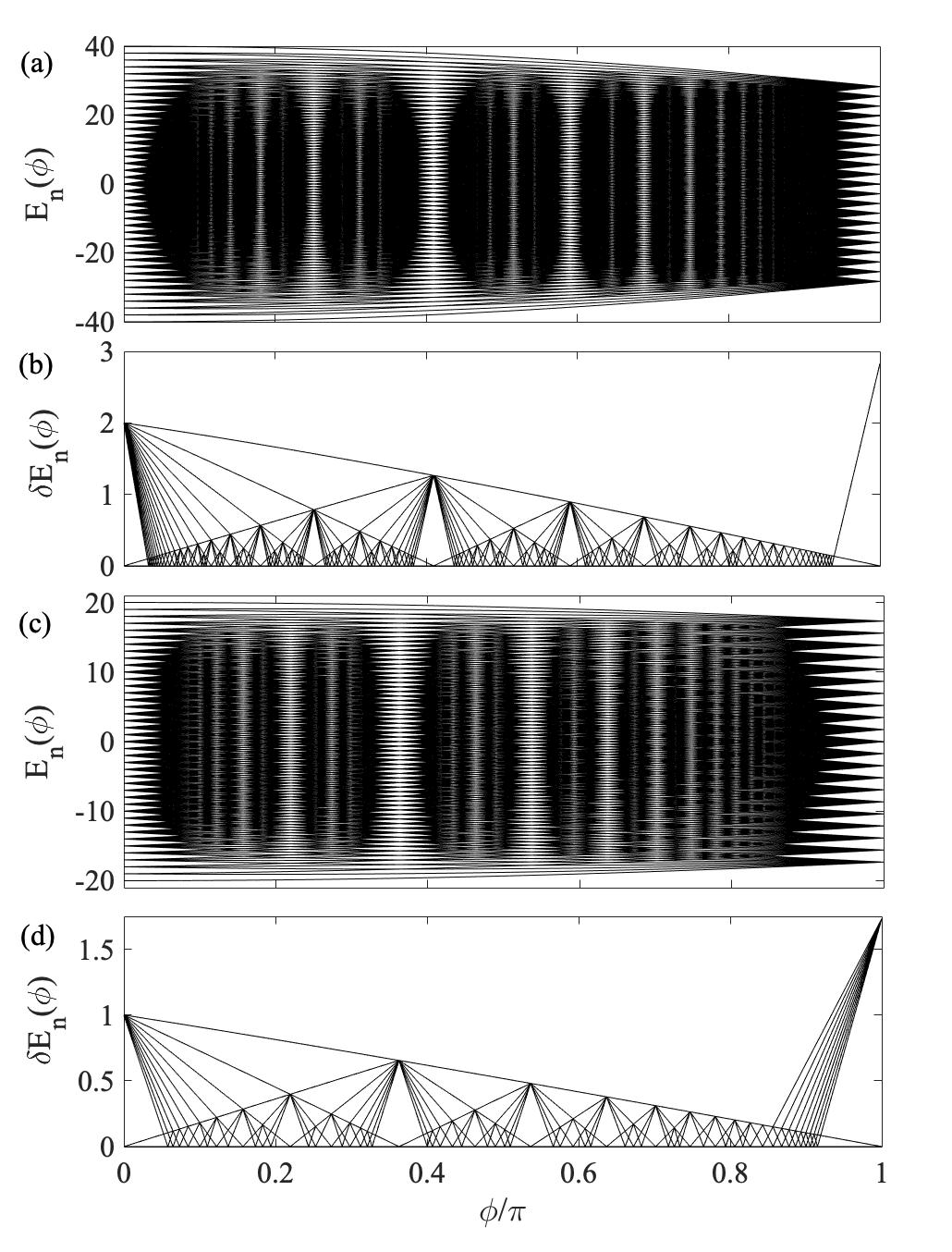}
\caption{The same as Fig.~\ref{fig:fig}, but for the four, upper two plots (a) and (b), and six mode, lower two plots (c) and (d), ring models. The number of bosons are $N=20$ and $N=10$ for the two cases respectively. By disregarding some of the terms in the Hamiltonian~(\ref{eq:phase}) to obtain the ring Hamiltonian~(\ref{ringham}), the number of conditions for having a quasi equidistant spectrum is reduced and it is possible to fulfill them simultaneously, and thereby the reappearance of fractal spectra for these two models. }
\label{fig:ring}
\end{figure}

The models defined via Eq.~(\ref{eq:phase}) are ``fully connected'', i.e. each mode can exchange excitations with every other mode. We may lift this by introducing selections among the tunneling processes, and in particular we consider bosonic chains with periodic boundary conditions
\begin{equation}\label{ringham}
\hat{H}=\sum_{i=1}^K \hat{a}^{\dagger}_{i}\hat{a}_{i+1}e^{i\phi/K}+\mathrm{h.c},
\end{equation}
where $\hat a_{K+1}=\hat a_1$. Since we impose periodic boundary conditions, and thereby the lattice has a ring structure, the phase $\phi$ becomes non-trivial in the sense that we cannot gauge it away. In the boson chain it mimics a magnetic flux through the lattice ring. For the fully connected four mode case, the FSL takes the form of a tetrahedron lattice. When we instead consider the four mode ring model, some of the links between sites in the tetrahedron lattice are cut; the number of sites each site couples to goes from 12 to eight. The resulting FSL has a rhombic structure, i.e. any 2D lattice plane forms a rhombic lattice. The characteristic equation for the corresponding matrix $h$ becomes $ \gamma^4-4\gamma^2+4\sin^2\left(\frac{\phi}{2}\right)=0$, and the four eigenvalues (roots) can be expressed as
\begin{equation}
    \gamma_{k\pm}=\pm\cos\left(\frac{\phi+\pi\cdot k}{2}\right), \qquad k\in\left\{0,1\right\}.
\end{equation}
This leads to the following condition for the flux $ \phi=2\arctan(q)$ with $q\in\mathcal{Q}$. Similarly, for the six-mode ring model we find the eigenvalues 
\begin{equation}
    \gamma_k=\pm 2\cos\left(\frac{\phi+2\pi\cdot k}{6}\right)\qquad k\in\left\{0,1,2\right\},
\end{equation}
giving the following condition for the quasi equidistant points $\phi=6\arctan\left(\frac{q}{\sqrt{3}}\right)$. In Fig.~\ref{fig:ring} we present the spectra and energy differences for the two ring models. Clearly, the two models' spectra display a similar fractal structure as for the three-mode model of Figs.~\ref{fig:fig} (a) and (b). Interestingly, the three-mode model is also a ring model just like those of Fig.~\ref{fig:ring}. However, we have numerically verified that the five- and eight-mode models do not reproduce fractal spectra, i.e. it is not a property of the ring-type coupling. We have noticed, though, that the spectrum $E_{\bf n}(\phi)$ is symmetric around $E=0$ for all fluxes $\phi$ whenever one considers an even number of boson modes. This suggest an additional chiral symmetry that holds only for ring models with an even number of modes.

\section{Evolution of initially localized states in the three-mode boson model}\label{sec:evolution}
To further study manifestations of the fractal spectrum, we turn to study the evolution over time. While our focus is on the three-mode case, let us start by a remark on the two-mode model, which produces a 1D FSL with tunneling rates $t\sqrt{(n+1)(N-n)}$. The Hamiltonian is easily diagonalized by a Bogoliubov transformation, but we use instead the Schwinger spin-boson mapping in which we identify a spin operator $\hat S_x=\frac{1}{2}\left(\hat a^\dagger\hat b+\hat b^\dagger\hat a\right)$, such that the Hamiltonian is simply $\hat H_\mathrm{int}=2\hat S_x$, with the total spin $S=N/2$. The spectrum is clearly equidistant, and any state returns to its initial state after the revival time $T_R=2\pi$. This model was studied by Christandl {\it et al.} in terms of perfect state transfer in 1D chains with spatially varying tunneling rates~\cite{cdel}, e.g. after half the revival time $T_R/2$ any initial state at one end of the chain will have traversed the chain to populate the other end. Recently the model was experimentally explored with the particular Hamiltonian realized by coupling two photonic modes via a beam splitter~\cite{FSsimulation}. To see the effect of the flux $\phi$ we need, however, to go up by one dimension.

As a demonstrating example we consider a system of $N=15$ bosons. Hence, the Fock states have a form $\ket{n_a, n_b, n_c}$ where $n_i\in \mathbb{N}$ and $n_a+n_b+n_c=15$. $N=15$ results in $\mathcal{S}=136$ different Fock states that make up the corresponding FSL. We shall look at the evolution of the two different initial states, taken to be $\ket{15,0,0}$ and $\ket{11,4,0}$, which both are initially completely localized on a single site located somewhere in the FSL. In order to characterize the extent of the states we consider the inverse partition ratio
\begin{equation}\label{ipr}
        IPR(t) =\sum_{\bf n}|\langle{\bf n}|\psi(t)\rangle|^4, 
\end{equation}
which gives the degree of localization in the FSL. For a fully localized state $IPR=1$, while for a maximally delocalized one $IPR=1/\mathcal{S}$, where $\mathcal{S}$ is, as before, the number of sites of the FSL. When it comes to the choice of fluxes we shall consider the time-evolution for the fluxes $\phi_1$ and $\phi_2$ according to Eq.~(\ref{eq:fract2}). In addition we pick a phase not coinciding with any equidistant point, namely $\tilde{\phi}=\pi/100$.

\begin{figure}[htb]
\centering
\includegraphics[width=8cm]{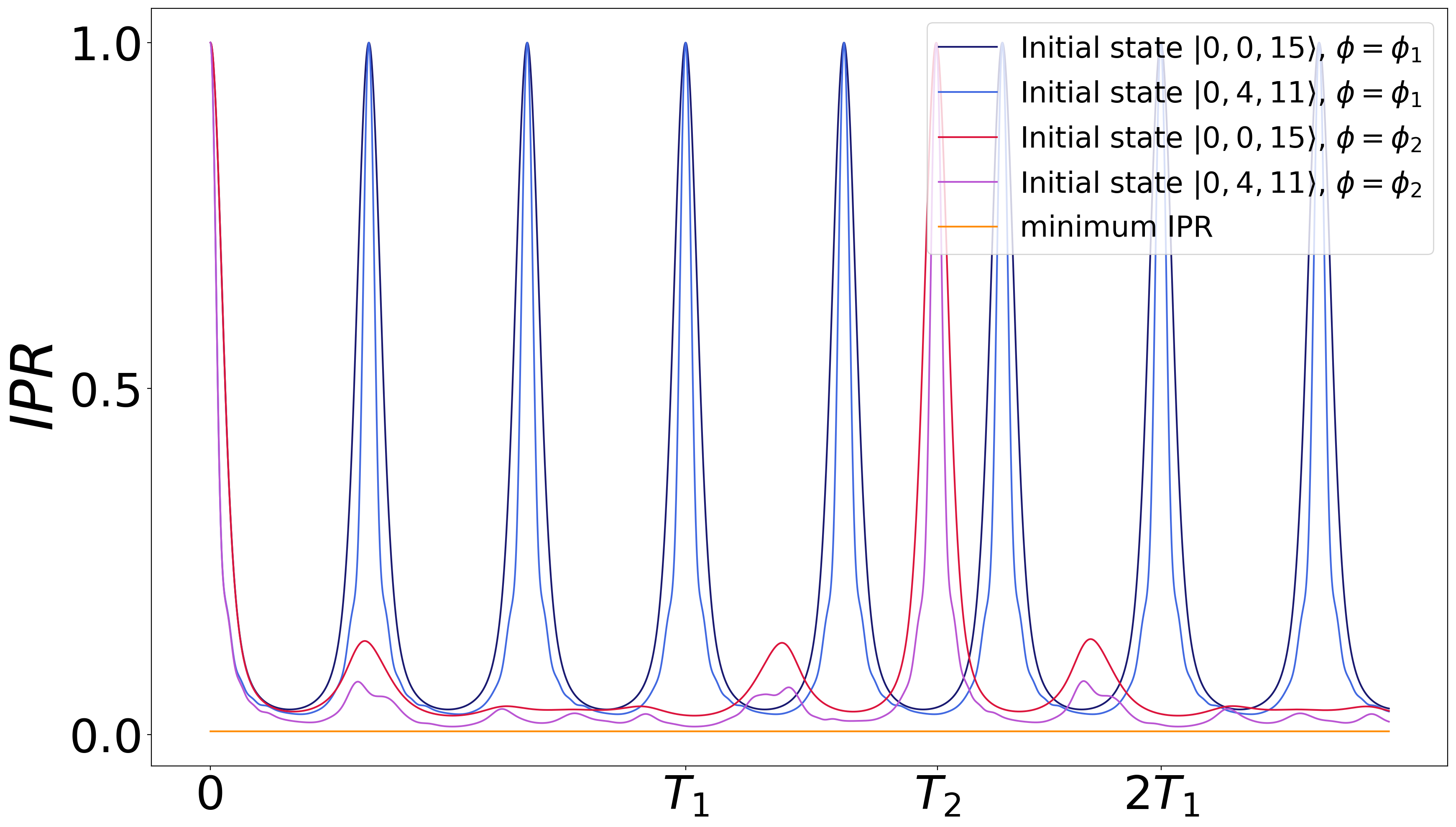}
\caption{(Color online) The $IPR$ for two different initial states, and two different fluxes corresponding to equidistant points. For both fluxes the states display perfect revivals, even though the revival times differ between the fluxes (but not between the states!). As a comparison, we also give the minimum possible $IPR$ (yellow line). How the $IPR$ is defined it is easy to get fooled that a small value necessarily implies an extended distribution. In fact, the distributions remain fairly localized throughout, see Fig.~\ref{fig:triangular}.}
\label{fig:Fock}
\end{figure}

\begin{widetext}
\begin{figure*}[htb]
\centering
\begin{subfigure}{}
\includegraphics[width=3.3cm]{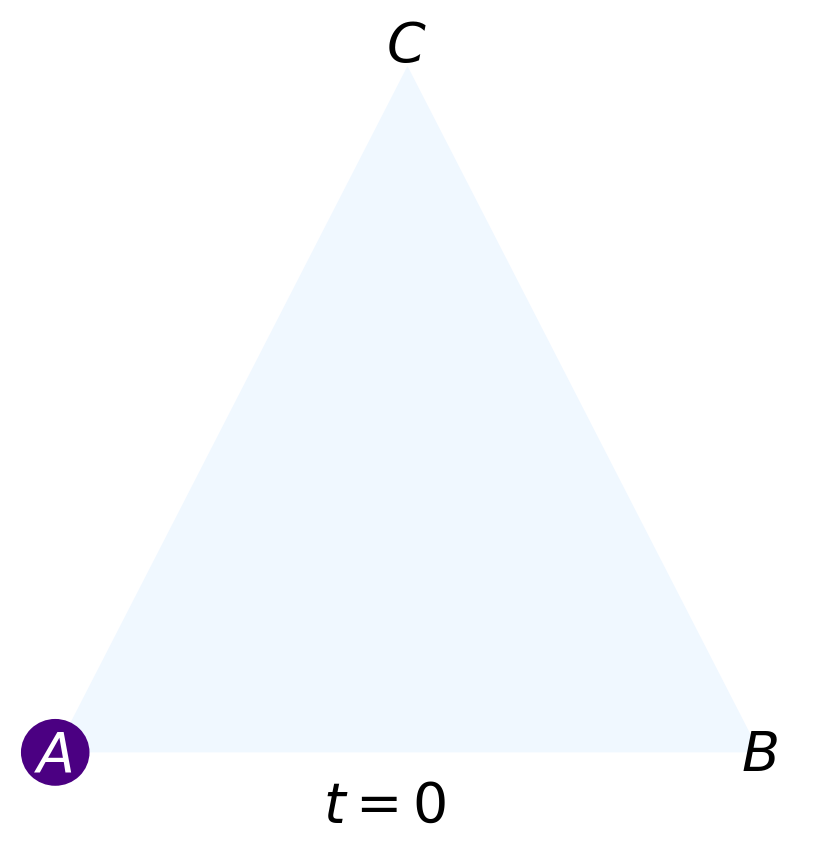}
\end{subfigure}
\begin{subfigure}{}
\includegraphics[width=3.3cm]{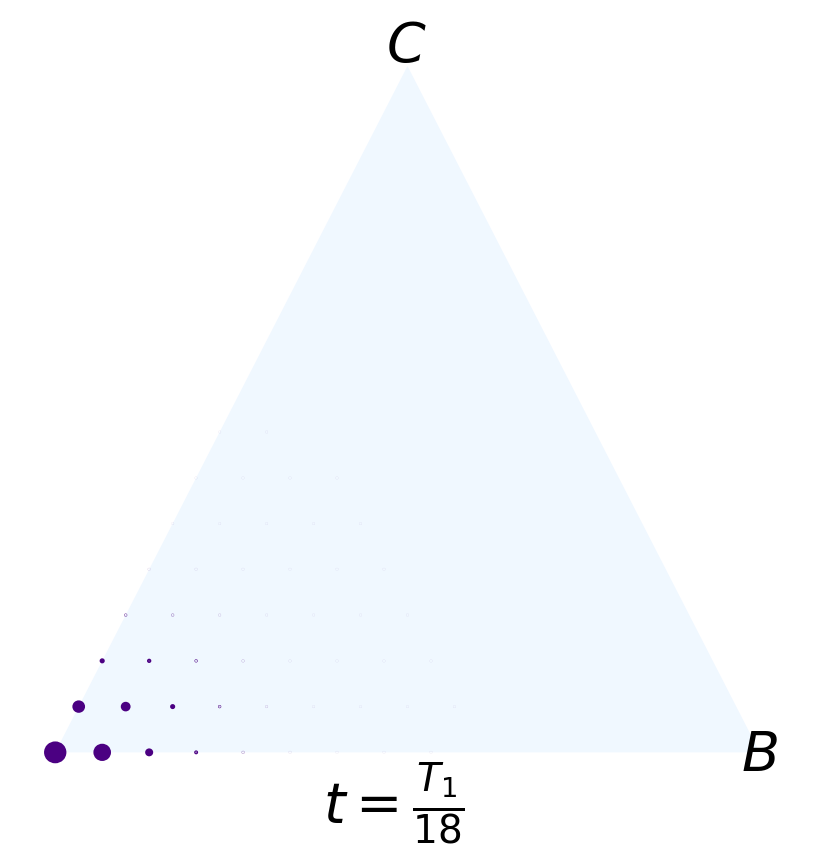}
\end{subfigure}
\begin{subfigure}{}
\includegraphics[width=3.3cm]{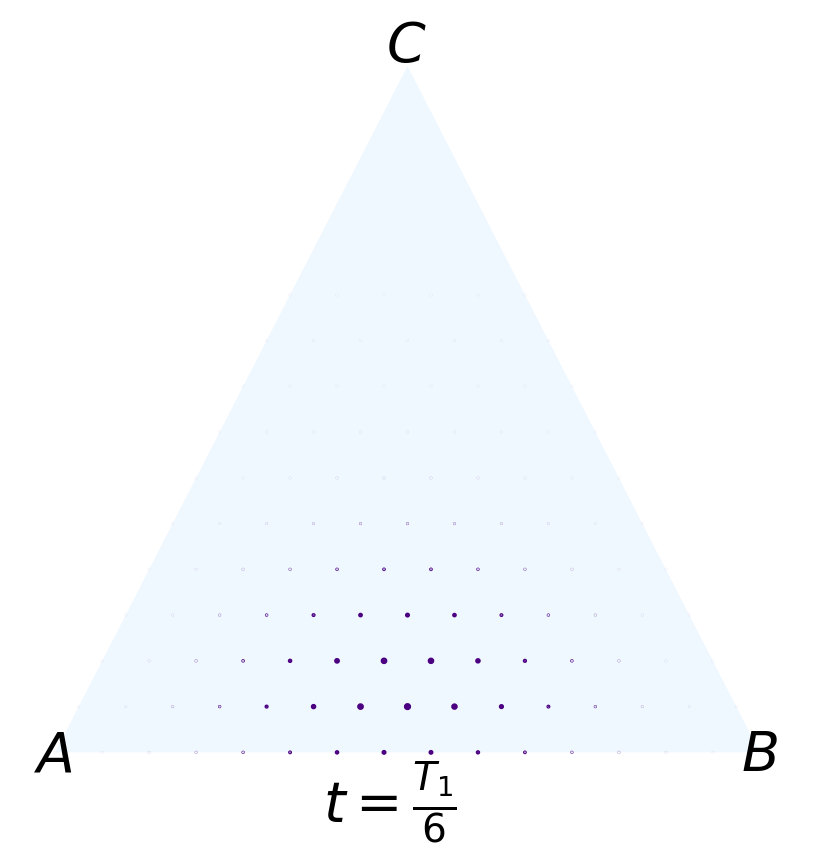}
\end{subfigure}
\begin{subfigure}{}
\includegraphics[width=3.3cm]{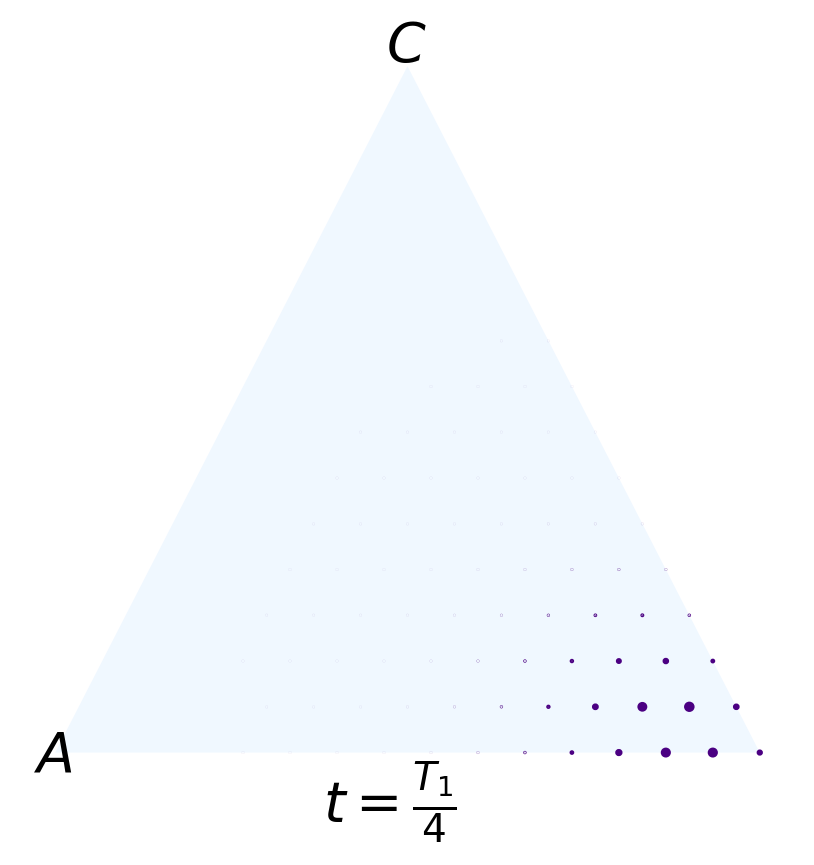}
\end{subfigure}
\begin{subfigure}{}
\includegraphics[width=3.3cm]{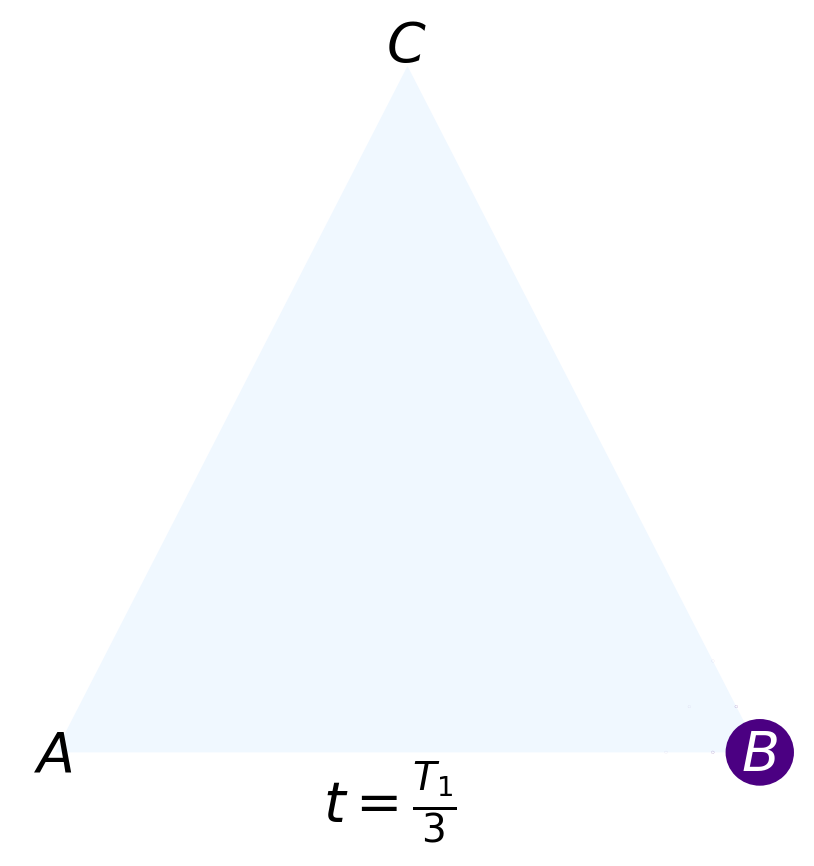}
\end{subfigure}
\begin{subfigure}{}
\includegraphics[width=18cm]{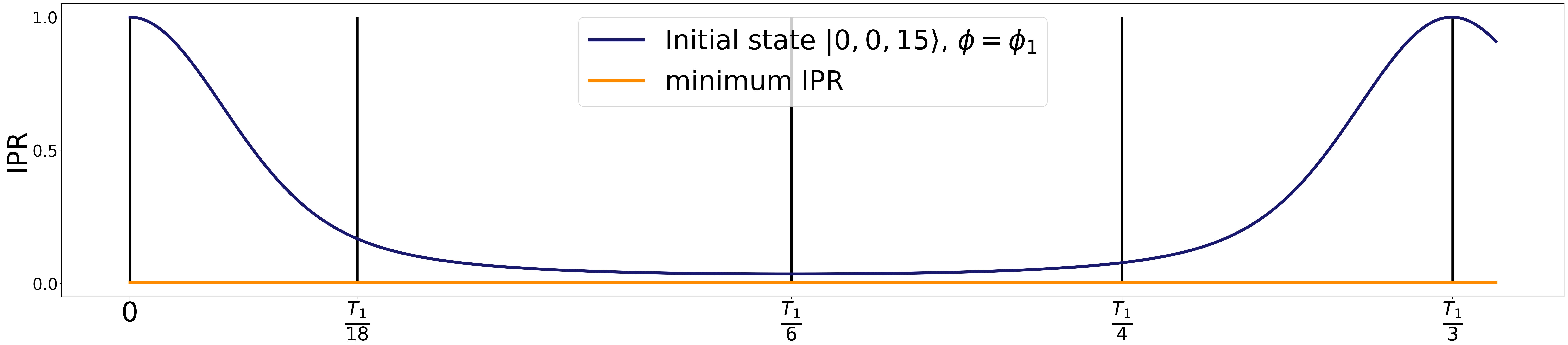}
\end{subfigure}
\caption{(Color online) The upper light blue triangles display snapshots of the Fock state distributions~(\ref{fockdist}) for the initial state $A=\ket{0,0,15}$ and for the flux $\phi_1=\pi/2$. Initially the distribution occupies a single corner site, and after $T_R/3$ it has been transferred to another corner site. In between it stays well localized and traverses along the lower edge of the lattice. The lower plot gives the corresponding $IPR$~(\ref{ipr}), with the vertical lines indicating the time instances used for the distribution snapshots. Note that for the center distribution at $T_1/6$ we have $IPR\ll1$, but still the distribution is well localized according to the upper figure.}
\label{fig:triangular}
\end{figure*}
\end{widetext}

The numerically extracted $IPR$s are depicted in Fig.~\ref{fig:Fock}. As long as the flux is taken to be one of the $\phi_j$ values we see perfect recurrences of the $IPR$ to its initial value $IPR=1$. The state, however, has not necessarily returned to its initial state at these instances, but it might have completely localized on another lattice site as will be clear below. In between the strong localization the state is spread over several lattice sites, but the $IPR$ never drops close to its minimum value (a small $IPR$ can still represent a localized distribution since it scales as $1/$occupied sites). Hence, even in-between the maximum localization the distribution remains fairly localized in the lattice, see further Fig.~\ref{fig:triangular}. For $\tilde{\phi}$ (not shown in the figure), when the spectrum is not quasi equidistant, we see no clear indicators of localization, even though the state is far from populating the whole lattice.

\begin{figure}[htb]
\centering
\includegraphics[width=8cm]{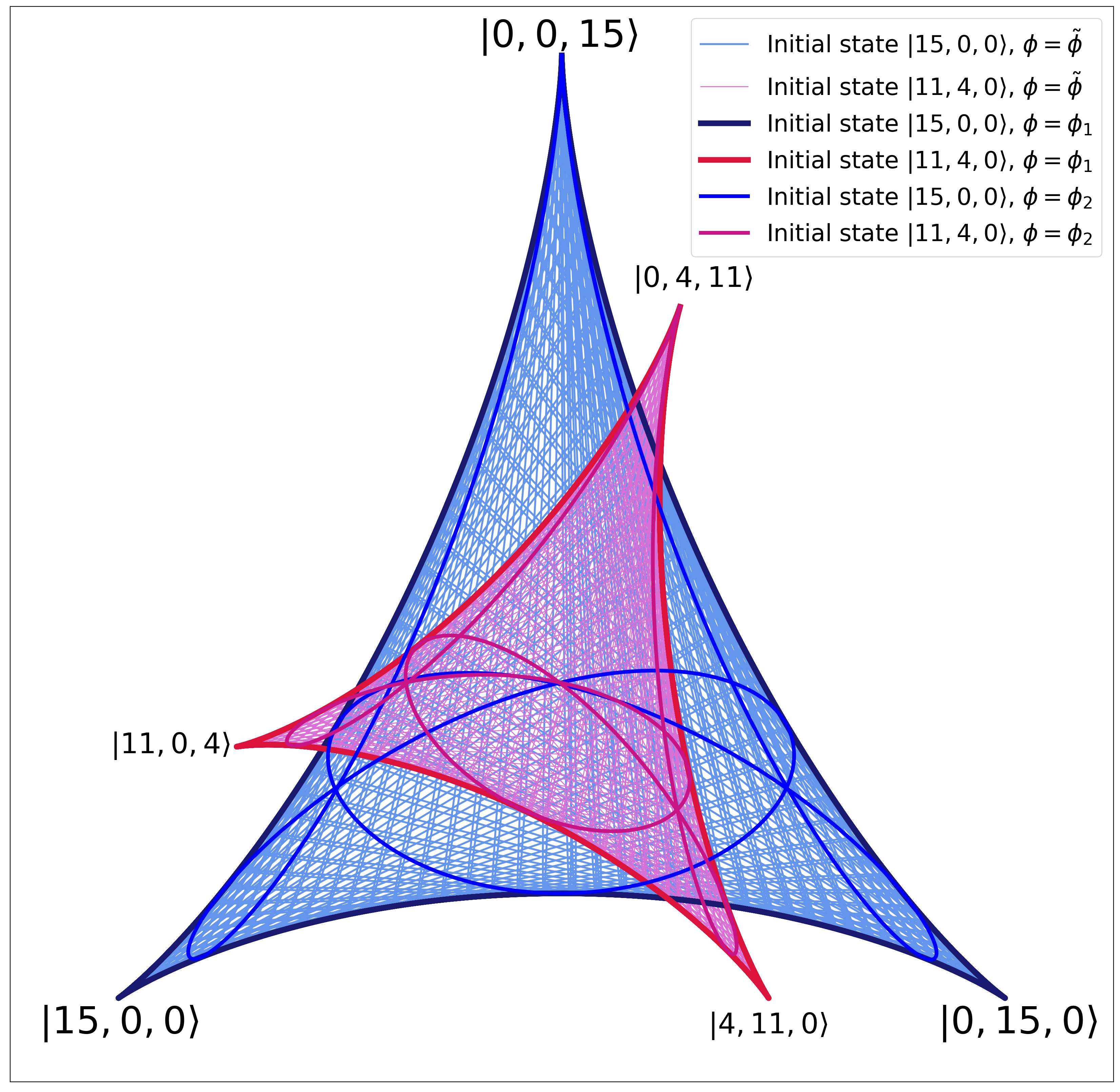}
\caption{(Color online) Examples of the evolution of the occupation vector $\vec{n}(t)=\left(n_a(t), n_b(t), n_c(t)\right)$, with $n_i(t)=\langle\hat n_i\rangle$ in the FSL for the three-mode model. The initial states are $\ket{0,0,15}$ (blue curves), and $\ket{0,4,11}$ (red curves), and the fluxes for the different curves are given in the inset. Note how the trajectories close whenever the flux is chosen as one of the equidistant points (signaling perfect revivals), while this is no longer true for a general flux (thin curves). In addition, the trajectories are found to be `shape invariant'; by which we mean that the two different initial states result in the same type of trajectories only rotated and scaled in size.}
\label{fig:trajectory}
\end{figure}

Recall that the constraint stemming from particle conservation removes one degree-of-freedom, which ensures, for example, that the participation vector $\vec{n}(\phi,t)=\left(n_a(\phi,t), n_b(\phi,t), n_c(\phi,t)\right)$, with $n_i(\phi,t)=\langle\hat n_i\rangle$, lives in a 2D plane, i.e. the FSL. Thus, the trajectory gives us an idea of how the full distributions move around in the lattice, even though it only captures the mean occupation but no higher moments of the distribution. In Fig.~\ref{fig:trajectory} we show examples of such trajectories for the same initial states and parameters as those used for Fig.~\ref{fig:Fock}. The first to notice is how the trajectory $\vec{n}(\phi_1,t)$ sets an outer boundary for all the other trajectories. 

%This implies that the following conditions
%\begin{equation}
%    max\left(n_i(\phi)\right)\leq max\left(n_j(\phi)\right) \quad\text{and} \quad \sum_i \langle n_i(\phi)\rangle \leq \langle n_i(\phi_1)
%\end{equation}
%must be fulfilled {\bf I don't understand these}. 
As pointed out, when $\phi_j$ agrees with one of the equidistant points we encounter perfect revivals at times 
\begin{equation}
    T_j = \frac{2\pi}{\Delta_j}
\end{equation}
where $\Delta_j$ is the aforementioned smallest energy splitting for a given $\phi_j$. The revivals manifest as closed trajectories of $\vec{n}(t)$ in the FSL. These loops form a symmetric pattern in the FSL, and the number of full 360$^\circ$ turns it makes before closing upon itself is found to be $2j-1$. Thus, for $j=1$ (i.e. $\phi_1=\pi/2$) the distribution makes a single loop around the lattice, which is found to follow closely the lattice edges, i.e. there always remains a large population imbalance among the three modes. For $\phi_2$, the distribution makes instead three loops before returning tho its initial states. For a flux away from any $\phi_j$ we do not encounter such revivals and hence no closing of the trajectories. For the closed loops there are `clustering points' in which many trajectories cross; the lattice origin being the most noticeable. A closer look reveals that for any initial Fock state, the resulting trajectories pass the origin (i.e. $n_a(t)=n_b(t)=n_c(t)=N/3$) for all $\phi_j$ with $j=2,3,5,6,8,9,\dots$. Upon comparing the trajectories originating from the two different initial Fock states, $\ket{0,0,15}$ vs. $\ket{0,4,11}$, we find that the shape and structure remain intact, however rotated and contracted. In the App.~\ref{sec:app} we derive semi-classical equations-of-motion that exactly reproduce the trajectories.  

The trajectories of Fig.~\ref{fig:trajectory} give the mean positions of the Fock state distribution
\begin{equation}\label{fockdist}
P({\bf n},t)=|\langle{\bf n}|\psi(t)\rangle|^2,
\end{equation}
but they say nothing about its actual shape. This is shown in Fig.~\ref{fig:triangular} by plotting the distribution $P({\bf n},t)$ at different times for the case of $\phi_1=\pi/2$. Clearly, the distribution remains fairly localized along the FSL edges throughout. Furthermore, it propagates counter-clockwise around the lattice which reflects the breaking of time-inversion as soon as the flux is non-zero (a zero flux does not produce a favoured direction for the propagating distribution). The propagation is reminiscent of a Hall current; a perpendicular magnetic field applied to a 2D electronic gas implies a current living on the edge of the system. In our case, the magnetic field is staggered, but we notice that the direction of the current can be reversed by flipping the sign of the fluxes.

\begin{figure}[htb]
\centering
\includegraphics[width=8cm]{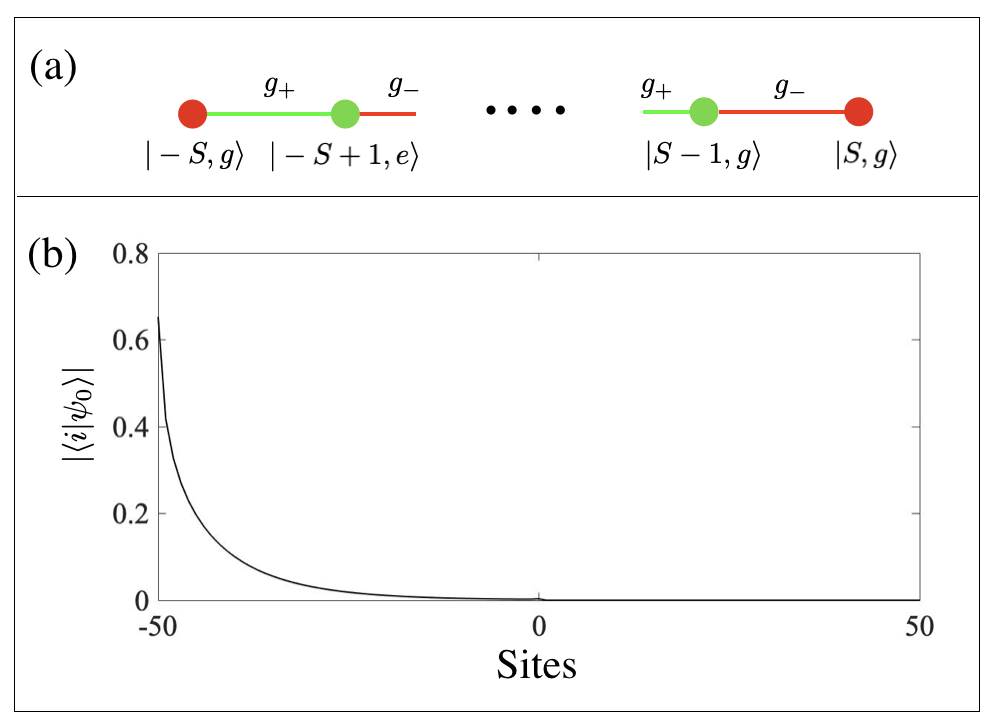}
\caption{(Color online) The FSL (a) emerging from the central spin model~(\ref{csm}) with anisotropic coupling amplitudes $g_\pm$. The red/greed dots mark Fock states with the spin-1/2 in the $|g\rangle/|e\rangle$-states respectively. The parity symmetry of the model implies a similar FSL but belonging to the other parity sector with the green/red dots swapped. In (b) we show the exponentially localized $E=0$ energy eigenestate $|\psi_0\rangle$ in an 101 site lattice with $g_+/g_-=0.9$. In the other parity sector there is an equivalent edge state localized to the right edge instead. }
\label{fig:ssh}
\end{figure}

\section{Towards the SSH model - and the appearance of edge states}\label{sec:ssh}
It is intriguing to speculate whether there is an underlying topological structure that causes the edge currents seen in Fig.~\ref{fig:triangular}. A more thorough discussion regarding this topic for the FSLs can be found in Refs.~\cite{jctop,FSsimulation}. Since translational invariance is absent we encounter discrete energies rather than bands, but it is interesting to note that for the chiral case with $\phi_1=\pi/2$ we find $Z=N/2+1$ or $Z=N/2+1/2$ degenerate $E=0$ energies for even and odd particle numbers $N$ respectively. These are not unique for the FSL, but appear also for the translationally invariant lattice with a staggered field. In other words, the degenerate $E=0$ energies in the FSL are the counterpart of a flat band. 

Instead of exploring properties of these $E=0$ states for the three-mode model, let us introduce another model also supporting $E=0$ eigenstates which are found to be living on the edge of the FSL. In order to construct a finite 1D FSL we do not a boson mode, but rather a spin-$S$ subsystem via the mapping
 \begin{equation}
     \begin{array}{ccc}
     \left(\hat a^\dagger+\hat a\right)& \rightarrow& \hat S_x,\\ 
     i\left(\hat a^\dagger-\hat a\right) & \rightarrow & \hat S_y,\\ 
     \hat a^\dagger\hat a& \rightarrow & \hat S_z,
     \end{array}
 \end{equation}
 where the $\hat{S}^\alpha$'s are the SU(2) spin operators for a spin-S particle. Contrary to the case of a boson mode, this introduces both an upper and a lower bound/edge for the lattice. When this is applied to the JC model one obtains a so called central spin model - which is a model where a central spin-$1/2$ particle interacts identically with $N$ non-interacting spin-$1/2$ particles which form the large spin-$S$ (i.e. $S=N/2$). The central spin model has served as a toy model for studying quantum criticality, as well as for analyzing non-Markovian decay of a qubit~\cite{centralspin}. To achieve the desired FSL we consider the central spin model on the form
 \begin{equation}\label{csm}
 \begin{array}{lll}
    \hat{H}_ {int}^{cs} & = &  g_x\hat{\sigma}^x\hat{S}^x +g_x\hat{\sigma}^y\hat{S}^y\\ \\
    & = & g_+\left(\hat{\sigma}^+\hat{S}^- + \hat{\sigma}^-\hat{S}^+\right)+g_-\left(\hat{\sigma}^+\hat{S}^+ + \hat{\sigma}^-\hat{S}^-\right),
    \end{array}
\end{equation}
where $g_{\pm}=g_x\pm g_y$. This is the spin analogue of the anisotropic quantum Rabi model~(\ref{anRabi}), i.e. the two types of interaction terms couple with different strengths; for $g_x=g_y$ we regain a JC type interaction, and for $g_y=0$ we instead get a quantum Rabi type interaction. Like for the quantum Rabi model, the central spin model supports a $\mathbb{Z}_2$-parity symmetry for general couplings $g_x$ and $g_y$. This results in two decoupled 1D chains, one for each parity. Furthermore, the chains have lengths $2S+1$, and provided $g_+\neq g_-$ and $g_+,\,g_-\neq0$, the tunneling rates will alternate between neighbouring sites (see Fig.~\ref{fig:ssh} (a)), similar to that of the SSH model. As for the SSH model, also for the central spin model one finds two $E=0$ eigenstates that become exponentially localized at either of the edges. More precisely, the probability to populate site $i$ for such a state scales as $P_i\propto\left(\frac{g_+}{g_-}\right)^{S-i}$. Each localized state belongs to either of the two parity sectors, and in Fig.~\ref{fig:ssh} (b) we show one of them for $g_+=0.9g_-$. For the two-mode JC model on resonance, one also find localized edge states~\cite{jctop}, but these, however, are not exponentially localized as the ones of the central spin model.

\begin{widetext}
\begin{table*}[htb]
  \centering
\begin{tabular}{|c|c|c|}
\hline
{\bf Model} & {\bf Interaction Hamiltonian} & {\bf Lattice type}  \\
  \hline \hline
Jaynes-Cummings & $\hat H_\mathrm{int}=g\left(\hat a^\dagger\hat{\sigma}_{-}+\hat a\hat{\sigma}_{+}\right)$ & Double-well, Fig.~\ref{fig:FSL} (a)  \\
  \hline
  Quantum Rabi & $\hat H_\mathrm{int}=g\left(\hat a^\dagger+\hat a\right)\hat\sigma_x$ & 1D chain, Fig.~\ref{fig:FSL} (b)   \\
  \hline
 Anisotropic quantum Rabi & $\hat H_\mathrm{int}=g_\mathrm{jc}\left(\hat{\sigma}_{+}\hat{a}+\hat{a}^\dagger\hat{\sigma}_{-}\right)+g_\mathrm{ajc}\left(\hat{\sigma}_{-}\hat{a}+\hat{a}^\dagger\hat{\sigma}_{+}\right)$ & Infinite SSH chain\\
  \hline
  $N$ atom Dicke & $\hat H_\mathrm{int}=g\left(\hat a^\dagger+\hat a\right)\hat S_x$ & Square lattice  \\
  \hline
  $N$ atom Tavis-Cummings & $\hat H_\mathrm{int}=g\left(\hat a^\dagger\hat S^-+\hat a\hat S^+\right)$ & $N$ potential-well  \\
   \hline
  Central spin model & $\hat H_\mathrm{int}=g_x\hat\sigma_x\hat S_x+g_y\hat\sigma_y\hat S_y$ & Finite SSH chain, Fig.~\ref{fig:ssh} (a)  \\
  \hline
  Driven quantum Rabi & $\hat H_\mathrm{int}=g\left(\hat a^\dagger+\hat a\right)\hat\sigma_x+\eta\left(\hat a^\dagger+\hat a\right)$ & Creutz ladder, Fig.~\ref{fig:FSL} (d)   \\
  \hline
  Single mode $\Lambda$ & $\hat H_\mathrm{int}=g\left(\hat a^\dagger+\hat a\right)\left(\hat\lambda^{(1)}+\hat\lambda^{(6)}\right)$ & Lieb ladder   \\
  \hline
  Two-mode JC & $\hat H_\mathrm{int}=g_a\left(\hat a^\dagger\hat{\sigma}_{-}+\hat a\hat{\sigma}_{+}\right)+g_b\left(\hat b^\dagger\hat{\sigma}_{-}+\hat b\hat{\sigma}_{+}\right)$ & SSH chain  \\
  \hline
  Two-mode detuned JC & $\hat H_\mathrm{int}=t\left(\hat a^\dagger\hat b+\hat b^\dagger\hat a\right)$ & CDEL chain  \\
  \hline
  Two-mode quantum Rabi & $\hat H_\mathrm{int}=g\left[\left(\hat a^\dagger+\hat a\right)+\left(\hat b^\dagger+\hat b\right)\right]\hat\sigma_x$ & (Layered) square lattice   \\
  \hline
   Two mode $\Lambda$ & $\hat H_\mathrm{int}=g_a\left(\hat a^\dagger+\hat a\right)\hat\sigma_{12}+g_b\left(\hat b^\dagger+\hat b\right)\hat\sigma_{23}$ & 2D Lieb lattice   \\
  \hline
  Three-mode JC & $\hat H_\mathrm{int}=g_a\left(\hat a^\dagger\hat{\sigma}_{-}+\hat a\hat{\sigma}_{+}\right)+g_b\left(\hat b^\dagger\hat{\sigma}_{-}+\hat b\hat{\sigma}_{+}\right)+g_c\left(\hat c^\dagger\hat{\sigma}_{-}+\hat c\hat{\sigma}_{+}\right)$ & Hexagonal lattice \\
  \hline
  Three-mode detuned JC & $\hat H_\mathrm{int}=t\left(\hat a^\dagger\hat b\, e^{i\varphi}+\hat b^\dagger\hat c+\hat a^\dagger\hat c+\text{h.c.}\right)$ & Triangular lattice, Fig.~\ref{fig:flux}   \\
  \hline
  Three-mode tripod & $\hat H_\mathrm{int}=g_a\left(\hat a^\dagger+\hat a\right)\hat\sigma_{12}+g_b\left(\hat b^\dagger+\hat b\right)\hat\sigma_{13}+g_c\left(\hat c^\dagger+\hat c\right)\hat\sigma_{14}$ & Perovskite lattice\\
  \hline
   Three-mode quantum Rabi & $\hat H_\mathrm{int}=g\left[\left(\hat a^\dagger+\hat a\right)+\left(\hat b^\dagger+\hat b\right)+\left(\hat c^\dagger+\hat c\right)\right]\hat\sigma_x$ & Cubic lattice   \\
  \hline
   Four-mode detuned JC1 & $\hat H_\mathrm{int}=t\sum_{j,i=1}^4e^{i\phi_{ij}}\hat a^\dagger_i\hat a_j$ & Tetrahedral lattice   \\
  \hline
   Four-mode detuned JC2 & $\hat H_\mathrm{int}=t\sum_{i=1}^4e^{i\phi_i}\hat a^\dagger_i\hat a_{i+1}+\text{h.c}.$ & Rhombic 3D lattice\\
  \hline
    \end{tabular}
  \caption{A collection of atom-light interaction Hamiltonians and their corresponding FSLs. Many of these appear frequently in the quantum optics community. Some of them we have discussed in some detail in the main text, e.g. the two-mode and three-mode JC models.    $\hat\sigma_{ij}=|i\rangle\langle j|+|j\rangle\langle i|$.}
  \label{focklat}
  \end{table*}
  \end{widetext}

\section{A zoo of Fock state lattices\label{sec:zoo}}

In quantum optics one encounters a range of models, comprising few degree-of-freedoms, that describe the interaction between quantized light and matter. Many of these are extensions and generalizations of the JC model towards multi-mode and multi-level atoms~\cite{Themis}. They often generate interesting FSLs with known, translationally invariant, counterparts in condensed matter physics. In Tab.~\ref{focklat} some of these models are listed, with their respective interaction Hamiltonians, and with information about the type of FSL that represents them. Below we discuss a few of these models
 
  \begin{itemize}
      \item {\it Driven quantum Rabi model.} The FSL belonging to the quantum Rabi model was shown in Fig.~\ref{fig:FSL} (b), where the $\mathbb{Z}_2$-parity symmetry splits the lattice into two 1D chains, one for each parity. We may break this parity by including a driving term, which induces tunneling along the legs of the FSL ladder, as depicted in Fig.~\ref{fig:FSL} (d). This is a special type of a Creutz ladder~\cite{Creutz}. The more standard Creutz ladder appears if we also include an atomic driving term that results in non-zero tunneling terms along the rungs of the ladder. In such a ladder we have competing nearest and next nearest neighbouring tunneling terms. A property of the Creutz ladder lattice is the presence of a dispersionless flat band when the tunneling rates are properly tuned. For our model, if we rotate the spin, $\hat\sigma_x\rightarrow\hat\sigma_z$, the Hamiltonian takes the simple form $\hat H_\mathrm{int}=(\eta\pm g)\left(\hat a^\dagger+\hat a\right)$. For $g=\eta$, i.e. the same amplitude of the tunneling rates, we find an infinite number of degenerate $E=0$ eigenstates. These are the corresponding flat band of the Creutz ladder, now manifested in the FSL. 
      
      \item {\it Two-mode $\Lambda$ model.} The Lieb lattice is another paradigm model hosting a flat band. In 2D it is a ``punctured'' squared lattice where every fourth lattice point has been excluded. This implies that the unit cell now contains three sites, with two of them having non-zero tunneling elements to two neighbours, and the third site couples to four neighbours. The lattice geometry provides a destructive interference mechanism that may completely hinder mobility, resulting in a $E=0$ flat band in the tight-binding limit~\cite{Lieb}. The remaining two bands touch in a Dirac cone in the center of the Brillouin zone. We can achieve this lattice type in a system of a three-level $\Lambda$ atom (two lower states and one excited), with the transition of each arm of the atom coupled to different boson modes. One finds that the destructive interference survives also in the FSL, and one thereby derives an infinite number of degenerate $E=0$ eigenstates. 
      
      \item {\it Three-mode tripod model.} The generalization of the three-level $\Lambda$ atom to four levels consists in adding one more arm, which is called the tripod atom. We can then couple each arm of the tripod to an individual mode such that we have a three-mode model. The resulting 3D lattice, bares similarities with the Lieb lattice, i.e. it is a punctured cubic lattice with a unit cell containing four points. The translational invaraint version is called a Perovskite type lattice, and as for the Lieb lattice it supports bands with non-trivial topology, i.e. their Chern numbers are non-zero~\cite{perov}. The flat bands in both the two-mode $\Lambda$ and three-mode tripod systems are related to the existence of dark states in these models, i.e. $E=0$ eigenstates. 
      
  \end{itemize}
  
  The list of Tab.~\ref{focklat} is by no means complete. As shown in Ref.~\cite{jctop}, many topological features of translationally invariant lattices survive in FSLs with the same geometry. Topological invariants, like the Chern number, are not directly applicable for these FSLs, and instead alternative topological numbers are studied~\cite{jctop}. Hence, given your favourite lattice with say some interesting properties, e.g. non-trivial topology, you can work backwards and ask for some interaction Hamiltonian that generates the corresponding FSL. For example, bi-layered lattices may host novel properties, like bi-layered Lieb lattices can produce high Chern numbers, and it is often straightforward to introduce more layers in the FSL by including additional internal atomic levels.

\section{Concluding remarks}\label{sec:conclusion}
In this work we have taken a new approach in order to analyze a range of different spin-boson models known from quantum optics. This idea relying on FSLs was first brought up in~\cite{FSL1}, and later extended in~\cite{jctop}. In these works, focus was on the multi-mode JC model. Part of the present work explores the same models, and we find a set of new properties. The spectra for the three-mode boson model, when a synthetic magnetic flux has been added, is found to be fractal, akin to the Hofstadter butterfly. The flux of the FSL is, however, staggered rather than uniform (a uniform flux would not render a fractal spectrum for the triangular FSL). The fractal structure cannot be explained by the same arguments used for the Hofstadter butterfly, as this relies on varying sizes of the lattice's unit cell. Instead, the emergence of our fractal structure is understood in terms of certain fluxes $\phi_j$ that causes a high degeneracy in the spectrum. For higher number of boson modes, the fractal structure is in general lost, but it can be restored for the four- and six-mode cases provided one adds some selection among the tunneling terms between the modes. 

The evolution of the system very much depends on the flux -- at the quasi equidistant points $\phi_j$ we have that the system shows perfect revivals. In the FSL, given that we start in a Fock states, for these fluxes the distribution $P({\bf n},t)$ remains well localized and follows closed loops. Since the flux breaks time-reversal symmetry, the loops are traversed either clockwise or anti-clockwise depending on the sign of $\phi_j$. One also finds that these closed trajectories have a mirror symmetry with respect to the axis intersecting the lattice origin and the site where the state is initialized. Furthermore, the index $j$ determines how many twists the distribution makes before it returns to the initial site. There is also an inherent scale invariance of these trajectories; for a given $\phi_j$ the trajectories are identical in structure irrespective of the specific initial Fock state, i.e. they only differ in orientation and size. 

We also demonstrated how fractal spectra appear in a four- and six-mode boson model. The properties of the corresponding trajectories, i.e. how the state evolves within the FSLs, was not consider, but we expect similar behaviour as for the three-mode model. The FSL of the three-mode model is two dimensional, the occupation vector $\vec n(t)$ evolves in a plane defined by a fixed total boson number $N$. Likewise, for the four-mode model we have a four component occupation vector $\vec n(t)$, but the conserved particle number implies that it is projected down to a 3D FSL.  

A boson degree-of-freedom results in infinite FSLs, unless some symmetry restricts it, e.g. particle conservation. Spin degrees-of-freedom have finite size Hilbert spaces and thereby finite size FSLs. We considered one such example, the anisotropic central spin model, and showed how this model has a topological property manifested in $E=0$ states exponentially localized to the edges of the lattice. Without going into any details we listed other models and mentioned which type of FSLs they generate. One interesting example being systems with ``flat bands'' being a huge $E=0$ degeneracy. Of course, one can consider other models beyond those mentioned in this paper. Another note we may make, here we use the Fock states, i.e. bare states, in order to construct the FSL. Shortly, the full Hamiltonian has the form $\hat H=\hat H_0+\hat H_\mathrm{int}$, where the Fock states are the eigenstates of $\hat H_0$, and $\hat H_\mathrm{int}$ constitute the FSL. We could, generalize this to a situation where $\hat H_0$ is not diagonal in the Fock basis, but instead in some other (dressed) basis. Take, for example, the driven JC model, $\hat H=\hat H_\mathrm{JC}+\eta\left(\hat a^\dagger+\hat a\right)$, and consider the JC dressed states, rather than the bare Fock states. The drive term will then constitute a new state space lattice (dressed state lattice), which in this example becomes a 1D chain. As another example, the quantum Rabi model could be written as $\hat H_\mathrm{R}=\hat H_\mathrm{JC}+g\left(\hat a^\dagger\hat\sigma^++\hat a\hat\sigma^-\right)$, such that it is expressed as a JC Hamiltonian plus the counter rotating terms~\cite{Themis}. In the JC dressed state basis, the resulting state space lattice formed by the counter rotating terms will be two decoupled 1D parity chains. Naturally, experimentally the bare Fock states are typically more relevant from a perspective of detection, but we mention this to stress the greater flexibility of the method by allowing for other bases.  
\appendix
\section{Equations of motion for the mean values of the three-mode boson model}\label{sec:app}
The multi-mode bosonic model~(\ref{eq:phase}) is quadratic and thereby solvable via a Bogoliubov transformation. Moreover, as a quadratic model it follows that an initial Gaussian state, e.g. any coherent or squeezed states, remains Gaussian at all times. Thus, a Gaussian quantum phase space distribution stay localized and follows the classical phase space trajectories. It is convenient to work with the classical canonical variables for position
\begin{equation}
\begin{split}
    \hat x=\frac{\left(\hat a^\dagger+\hat a\right)}{\sqrt{2}}
\end{split}
\end{equation}     
and momentum
\begin{equation}
\begin{split}
    \hat p_x=\frac{\left(\hat a^\dagger-\hat a\right)}{\sqrt{2}}.
\end{split}
\end{equation}
Using that the Heisenberg and Hamilton's equations have the same structure for a quadratic bosonic Hamiltonian, for the three-mode case we get the equations of motion 
\begin{equation}
\begin{array}{lll}
\dot x_a & = & -p_b-p_c\\
\dot x_b & = & -p_a-\cos(\varphi)p_c-\sin(\varphi)x_c,\\
\dot x_c & = & -p_a-\cos(\varphi)p_b+\sin(\varphi)x_b,
\end{array}
\end{equation}
and
\begin{equation}
\begin{array}{lll}
\dot p_a & = & x_b+x_c,\\
\dot p_b & = & x_a+\cos(\varphi)x_c-\sin(\varphi)p_c,\\
\dot p_c & = & x_a+\cos(\varphi)x_b+\sin(\varphi)p_b,
\end{array}
\end{equation}
such that $\langle\hat n_\alpha\rangle_t=\left(p_\alpha^2(t)+x_\alpha^2(t)\right)/2$. Since a Gaussian state (apart from the vacuum) does not contain a fixed number of bosons $N$, it means for the FSLs that it will populate different triangular FSLs -- one for each boson number $N$. However, the shape and structure of the trajectories in the FSLs are invariant for different $N$-values; the particle number only provides a scaling of the trajectories as demonstrated in Fig.~\ref{fig:trajectory}. 

For the above equations to reproduce the correct trajectories of figs.~\ref{fig:trajectory}, where the initial state is a Fock state $|N,0,0\rangle$, one should use the initial conditions $(x_a,x_b,x_c,p_a,p_b,p_c)=(\sqrt{2N},0,0,0,0,0)$. We have numerically conformed that these trajectories agree with those obtained from diagonalizing the full $N$-boson Hamiltonian. 

\begin{acknowledgements}\label{sec:acknowledgements}
The authors thank Themistoklis Mavrogordatos for helpful feedback on the manuscript. We acknowledge financial support from VR-Vetenskapsr\aa set (The Swedish Research Council), and KAW (The Knut and Alice Wallenberg foundation).
\end{acknowledgements}

\end{document}